\documentclass[pdflatex,sn-mathphys-num]{sn-jnl}

\usepackage{graphicx}%
\usepackage{multirow}%
\usepackage{amsmath,amssymb,amsfonts}%
\usepackage{amsthm}%
\usepackage{mathrsfs}%
\usepackage[title]{appendix}%
\usepackage{xcolor}%
\usepackage{textcomp}%
\usepackage{manyfoot}%
\usepackage{booktabs}%
\usepackage{algorithm}%
\usepackage{algorithmicx}%
\usepackage{algpseudocode}%
\usepackage{listings}%
\usepackage{booktabs}
\usepackage{threeparttable}
\usepackage{multirow}
\usepackage{adjustbox}

\theoremstyle{thmstyleone}%
\theoremstyle{thmstyletwo}%

\theoremstyle{thmstylethree}%

\unnumbered

\begin{document}

\title[Synthesizability Prediction of Crystalline Structures with Structure-Aware Feature Learning and Uncertainty Quantification]{Synthesizability Prediction of Crystalline Structures with Structure-Aware Feature Learning and Uncertainty Quantification}


\author[1]{\fnm{Danial} \sur{Ebrahimzadeh}}\email{danial.ebrahimzadeh@ou.edu}

\author[1]{\fnm{Sarah} \sur{Sharif}}\email{s.sh@ou.edu}

\author*[1]{\fnm{Yaser Mike} \sur{Banad}}\email{bana@ou.edu}

\affil*[1]{\orgdiv{School of Electrical and Computer Engineering}, \orgname{University of Oklahoma}, \orgaddress{\street{110 W. Boyd St.}, \city{Norman}, \postcode{73019}, \state{Oklahoma}, \country{USA}}}


\abstract{Predicting which hypothetical inorganic crystals can be experimentally realized remains a central challenge in accelerating materials discovery. SyntheFormer is a positive-unlabeled framework that learns synthesizability directly from crystal structure, combining Fourier-transformed crystal properties (FTCP) representation with structure-aware feature extraction, Random-Forest feature selection, and a compact deep MLP classifier. The model is trained on historical data from 2011 through 2018 and evaluated prospectively on future years from 2019 to 2025, where the positive class constitutes only 1.02 percent of samples. Under this temporally separated evaluation, SyntheFormer achieves a test AUC of 0.735, AUPRC of 0.099 and 97.6 percent recall at 94.2 percent coverage with dual-threshold calibration. Direct prospective validation supports this result as two materials that were unlabeled at the time of data curation, Y$_6$Fe(SiS$_7$)$_2$ and BaYb$_2$F$_8$, were assigned high SyntheFormer scores (0.961 and 0.753, respectively) and were subsequently reported experimentally. Crucially, the model recovers experimentally confirmed metastable compounds that lie far from the convex hull and simultaneously assigns low scores to many thermodynamically stable yet unsynthesized candidates, demonstrating that stability alone is insufficient to predict experimental attainability.}




\maketitle

\section{Introduction}\label{sec1}

\begin{figure*}[!t]
	\centering
	\includegraphics[scale=0.98, height=0.40\textheight, width=0.99\textwidth, trim=18cm 36cm 25cm 25cm, clip]{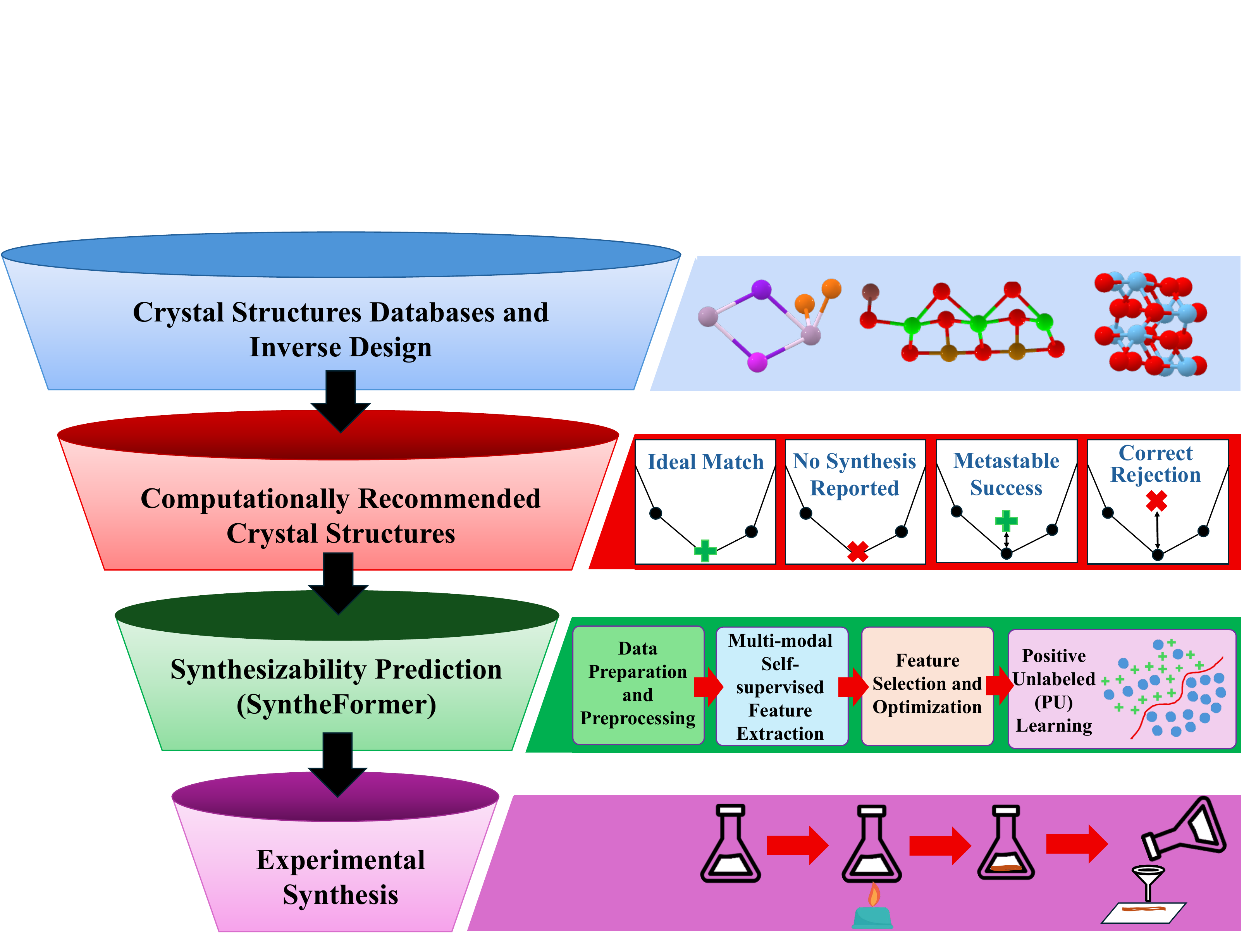}
	\caption{SyntheFormer integration in computational materials discovery workflow. The framework positions synthesizability prediction as a critical filtering step between computational crystal structure generation and experimental synthesis. Candidate structures sourced from crystal structure databases or inverse design tools are evaluated through a four-stage pipeline comprising data preparation and preprocessing, multi-modal self-supervised feature extraction, feature selection and optimization, and positive-unlabeled (PU) learning. The model assigns each candidate to one of four outcome categories: ideal match (high-scoring confirmed synthesizable), no synthesis reported (low-scoring unlabeled), metastable success (high-scoring despite thermodynamic instability), or correct rejection (low-scoring non-synthesizable). \label{fig1}
	}
\end{figure*}

Discovering synthesizable inorganic crystalline materials remains a grand challenge in materials science \cite{bartel2022review}. Despite centuries of exploratory synthesis, only on the order of $10^5$--$10^6$ \cite{pearson1901liii,kajita2017universal,korolev2020machine,zagorac2019recent} distinct inorganic compounds have been experimentally realized, out of an estimated $\sim$$10^{10}$ \cite{xie2018crystal} theoretically possible combinations \cite{qiu2025massive,park2024has}. Accelerating the expansion of this known chemical space is essential to enable breakthroughs in energy storage, electronics, quantum technologies, and other emerging technologies \cite{qiu2025massive}. High-throughput \textit{in silico} screening methods (e.g. density-functional theory \cite{kohn1965self,jain2013commentary,kirklin2015open,curtarolo2012aflow}) can propose many thermodynamically stable candidate structures \cite{qiu2025massive}, but stability alone is not a reliable proxy for synthesizability \cite{sun2016thermodynamic,lee2022machine,yu2024far}. Numerous hypothetical compounds predicted to lie on a convex hull remain unrealized due to kinetic barriers, while conversely some experimentally known crystals are metastable relative to their phase diagrams yet can be synthesized under specialized conditions \cite{kim2025explainable,song2025accurate,sokolikova2020direct}. Thus, proximity to the ground state is neither necessary nor sufficient for attainability \cite{sun2016thermodynamic}, and high-energy phases may be synthesized given the right conditions \cite{ye2018deep}. This gap has motivated data-driven approaches that learn from empirical record of successful syntheses to predict practical accessibility \cite{davariashtiyani2021predicting,kononova2019text,cruse2023text}. Incorporating synthesizability prediction as a filtering step in computational materials discovery workflows represents a paradigm shift from traditional approaches that rely solely on thermodynamic stability (Fig. \ref{fig1}).

Notably, the nature of available data makes synthesizability prediction a highly imbalanced positive-unlabeled (PU) classification problem \cite{bekker2020learning,kiryo2017positive,elkan2008learning,chung2025solid}. While crystallographic databases such as ICSD \cite{levin2018NIST} and the Materials Project \cite{jain2013commentary} contain thousands of confirmed synthesizable compounds, the space of chemically reasonable but unexplored candidates is vastly larger. We, therefore, have abundant confirmed positives but virtually no confirmed negatives. The millions of potential compounds absent from these databases may be genuinely unsynthesizable or simply undiscovered \cite{kim2025explainable}. This ambiguity imposes unique challenges for machine learning models, which must carefully handle unlabeled data and mitigate bias in order to generate meaningful predictions of synthesizability from limited ground truth \cite{antoniuk2023predicting}.

In recent years, machine learning methods have emerged to tackle this challenge by exploiting the patterns embedded in known materials data  \cite{davariashtiyani2021predicting,zhu2023predicting}. Broadly, prior works can be divided into composition-based and structure-based models, each with distinct strengths and limitations \cite{xie2018crystal}. Composition-based methods learn directly from stoichiometric formulas, enabling efficient screening across broad chemical spaces. For example, Jang et al. developed a semi-supervised positive-unlabeled classifier for binary-to-quaternary compositions that successfully identified formulas likely to correspond to synthesizable compounds \cite{jang2024synthesizability}. Antoniuk et al. introduced SynthNN, a deep learning model trained on the entire set of known inorganic compositions, which significantly outperformed convex-hull stability filters in precision \cite{antoniuk2023predicting}. Similarly, Zhu et al. applied machine learning over large materials databases to rank the synthesizability of candidate compositions, producing prioritized candidates for experimental validation \cite{zhu2023predicting}. While composition-based models excel at scalability, a fundamental drawback is that they cannot distinguish among polymorphs or capture structural subtleties influencing synthesizability, since only elemental makeup is considered \cite{kim2025explainable}. A controlled comparison isolating the effect of representation confirms this limitation quantitatively, with composition-only baselines achieving 2.7 to 3.3 times lower AUPRC than the structure-aware FTCP representation under identical model architecture and training conditions.

In contrast, structure-based models incorporate explicit information about the atomic configuration and bonding network, allowing them to distinguish stable and unstable polymorphs and capture specific structural motifs that influence synthesizability. Jang et al. pioneered a partially supervised learning approach using crystal structure descriptors, such as atomic environment fingerprints, to classify whether a given crystal structure is synthesizable or not \cite{jang2020structure}. Davariashtiyani et al. instead transformed crystal structures into 3D voxel grids with atomic positions encoded as densities, training convolutional neural networks to identify structural anomalies linked to non-synthesizable crystals across a broad chemical space \cite{davariashtiyani2021predicting}. Graph neural networks have also been employed to perovskite families, where learned representations of structural connectivity achieved accurate predictions of synthesizable outcomes \cite{gu2022perovskite}. SchNet-based co-training approaches \cite{amariamir2025syncotrain} and bootstrap PU-CGCNN \cite{jang2020structure} ensembles have extended structure-based synthesizability prediction to broader chemical spaces, yet their performance under realistic forward-looking temporal evaluation remains uncharacterized. These approaches highlight the value of structure-informed modeling, yet they remain limited to cases where candidate crystal structures (not just composition) can be generated or hypothesized in advance. Moreover, like composition-based models, they must contend with extreme data imbalance, often requiring specialized PU-learning techniques---such as bagging-based \cite{mordelet2014bagging} ensemble undersampling with calibrated probability thresholds---to avoid biased decision boundaries between ``synthesizable'' and ``unsynthesizable'' materials \cite{bekker2020learning}. Furthermore, atomic-graph representations encode only local coordination environments within a finite cutoff radius (typically 5 to 8 Å) and therefore cannot directly capture the long-range periodicity and reciprocal-space features that are central to diffraction-based characterization of crystal structures. A controlled comparison under identical training conditions confirms this limitation, with standard CGCNN \cite{xie2018crystal} and MEGNet \cite{chen2019graph} architectures achieving test AUPRC more than 3 times lower than the FTCP-based pipeline.

To overcome the limitations of composition- and structure-based approaches, we introduce SyntheFormer, a framework that unifies both perspectives through multi-pathway attention mechanisms applied to the Fourier-transformed crystal properties (FTCP) representation \cite{ren2022invertible,du2024ctgnn,vaswani2017attention}. SyntheFormer integrates elemental composition, real-space structural descriptors, and reciprocal-space features into a single crystallographic fingerprint, analogous to combining local atomic arrangements with diffraction-like signatures \cite{qiu2025massive,ziletti2018insightful}. This unified representation enables the model to capture both broad chemical trends and subtle structural factors that govern synthesizability, positioning synthesizability prediction as a critical filtering step between computational crystal generation and experimental synthesis (Fig. \ref{fig1}).

The FTCP representation partitions crystallographic information into six complementary components: elemental composition, lattice parameters, atomic sites, site occupancy, reciprocal space features, and structure factors, allowing systematic analysis of diverse crystallographic aspects through specialized attention-based pathways. SyntheFormer's multi-stage architecture processes these components with transformer modules: multi-head attention \cite{taniai2024crystalformer} to model spatial relationships among atomic sites, while cross attention and graph attention analyze occupancy patterns and structural connectivity. The framework combines these attention-based features via self-supervised learning to mitigate temporal distribution shifts, enabling the model to learn robust crystallographic representations independent of synthesis labels \cite{kim2024predicting}. Finally, SyntheFormer incorporates adaptive threshold strategies \cite{sahoo2021reliable} that reformulate positive-unlabeled learning into a practical screening tool, providing multi-level confidence assessment that balances recall, precision, and uncertainty in real-world materials discovery.

SyntheFormer demonstrates robust temporal generalization under a realistic forward-looking evaluation protocol. When trained on materials reported from 2011 to 2018 and tested on future candidates from 2019 to 2025, where the positive rate drops to 1.02\%, the model maintains a test AUC of 0.735 and AUPRC of 0.099. Beyond ranking performance, SyntheFormer incorporates validation-calibrated dual- and triple-threshold decision strategies that convert the classifier into a practical screening tool for materials discovery. This design emphasizes high-recall candidate prioritization while explicitly accounting for uncertainty, reducing the risk of overlooking experimentally attainable compounds in extremely imbalanced prospective settings.

By identifying synthesizable candidates across diverse chemical spaces while providing explicit uncertainty quantification, SyntheFormer represents a significant advancement in computational materials discovery. The approach offers a practical tool for accelerating experimental synthesis efforts by directing resources toward the most promising candidates within the vast landscape of theoretically possible materials. Beyond its immediate utility, SyntheFormer establishes a generalizable paradigm for addressing positive-unlabeled learning challenges in materials science, where negative data are inherently scarce. This positions SyntheFormer not only as a predictor of synthesizability, but also as a foundation for future integration with inverse design pipelines \cite{ebrahimzadeh2025accelerated}, autonomous synthesis laboratories, and closed-loop discovery systems aimed at bridging computation and experiment.

\section{Results}\label{sec2}
\subsection{Data for synthesizability prediction}\label{subsec2_1}

\begin{figure*}[!t]
	\centering
	\includegraphics[scale=1, height=0.68\textheight, width=0.99\textwidth, trim=8cm 0cm 25cm 0cm, clip]{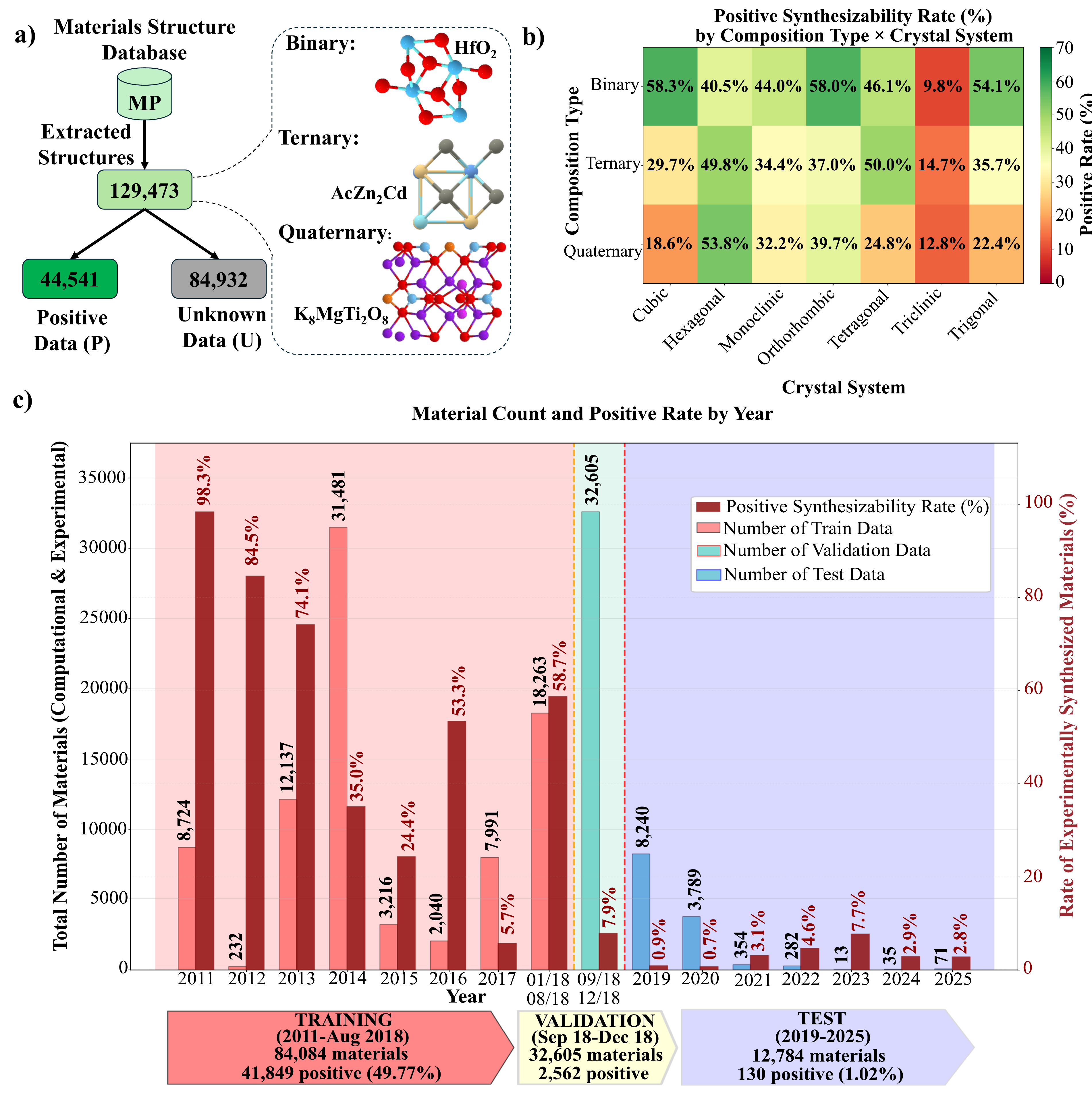}
	\caption{Dataset composition and temporal split (N = 129,473 inorganic crystals). \textbf{a} Materials Project dataset with 44,541 ICSD-confirmed positives and 84,932 unlabeled candidates, shown with representative binary, ternary, and quaternary structures. \textbf{b} Heatmap of positive synthesizability rate (\%, defined as the fraction of ICSD-confirmed entries) across composition type and crystal system. \textbf{c} Temporal splitting into training (2011--Aug 2018, N = 84,084, 49.8\% positive), validation (Sep--Dec 2018, N = 32,605, 7.9\% positive), and test (2019--2025, N = 12,784, 1.02\% positive) sets. This chronological split mimics real-world deployment, ensuring models are trained on past data and evaluated on future discoveries. \label{fig2}
	}
\end{figure*}

Our study leverages a dataset of inorganic crystalline materials from the Materials Project \cite{jain2013commentary}, spanning 2011--2025 and encompassing binary, ternary, and quaternary compositions. Positive labels correspond exclusively to entries cross-referenced with the Inorganic Crystal Structure Database (ICSD) \cite{levin2018NIST}, ensuring that only experimentally synthesized compounds are marked as such, while all others are treated as unlabeled candidates, representing plausible but as-yet unconfirmed structures. The dataset comprises 129,473 crystal structures, of which 44,541 (34.4\%) are positive and 84,932 (65.6\%) remain unlabeled (Fig. \ref{fig2}a). This imbalance highlights a central challenge in synthesizability prediction: as compositional complexity increases, the chemical space expands rapidly while the likelihood of experimental realization declines correspondingly.

The overall composition distribution shows that ternaries are most numerous (also within train and validation data), followed by quaternaries and binaries; however, the synthesis rate is highest for binaries (47.9\%), compared with ternaries (35.6\%) and quaternaries (26.3\%) (Figs. S1--S2). The dataset further reveals systematic variations across crystallographic symmetry classes as shown in Fig. \ref{fig2}b), that across all materials, the Orthorhombic system is the most frequent at 22.4\%, closely followed by Monoclinic at 22.2\% (Fig. S6). Within each composition type, the two highest-performing categories are binary cubic (58.3\%) and binary Orthorhombic (58.0\%), whereas the lowest are binary triclinic (9.8\%) and quaternary triclinic (12.8\%). These results underscore that compositional complexity strongly reduces the likelihood of successful synthesis. The most common elements are overwhelmingly Oxygen (O) across all composition types, with its frequency spiking to 34,569 in quaternary compounds. Further inspection of specific compositions reveals $\mathrm{Li_7 Mn_2 Co_5 O_{16}}$ as the most frequent quaternary compound, and $\mathrm{Si O_2}$ as the most frequent binary compound, reflecting a strong emphasis on Li-ion battery and geological materials, respectively (Figs. S7--S8). These observations highlight how both compositional complexity and crystal symmetry shape the distribution of synthesizable structures within our dataset.

To mimic the real-world task of assessing synthesizability of future candidates, we employed a temporal splitting strategy (Fig. \ref{fig2}c). All structures reported from 2011--August 2018 (84,084 entries, 49.8\% positive) were used for training, while a four-month buffer period (September--December 2018, 32,605 entries, 7.9\% positive) was reserved for validation. The held-out test set consists of 12,784 materials reported from 2019--2025, where only 1.02\% are experimentally confirmed (Table S1). Temporally, the rate of discovering synthesizable materials has generally declined, showing a linear trend of -6.38\% per year as shown in Fig. S3. This steep drop in positives reflects the growing reliance on computational predictions and the shift toward more complex chemistries in recent years. Such an extreme imbalance transforms the test phase into a particularly demanding benchmark: models must identify a handful of true synthesizable compounds hidden among thousands of unlabeled entries. By design, this setup mirrors deployment conditions where false negatives (overlooking synthesizable candidates) represent lost discovery opportunities. The pronounced skew between training and test distributions underscores the need for representations and learning strategies tailored to rare-event prediction, a challenge addressed in the following subsections.

\subsection{Data Representation and Feature Extraction}\label{subsec2_2}

\begin{figure*}[!t]
	\centering
	
	\includegraphics[scale=1, height=0.48\textheight, width=0.98\textwidth, trim=0cm 0cm 0cm 0cm, clip]{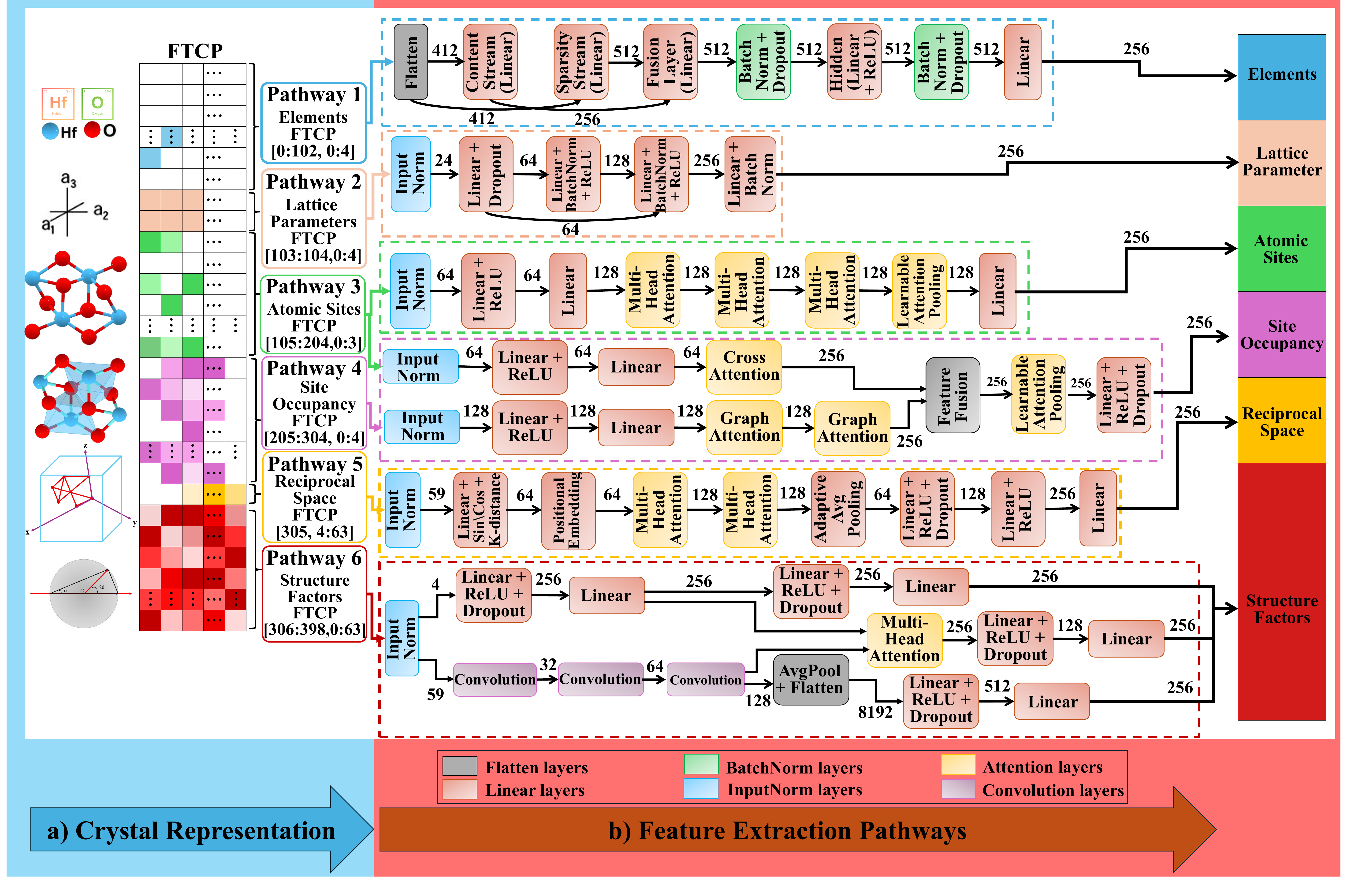}
	
	\vspace{0.01cm}
	
	\includegraphics[scale=1, height=0.20\textheight, width=0.98\textwidth, trim=0cm 30cm 0cm 0cm, clip]{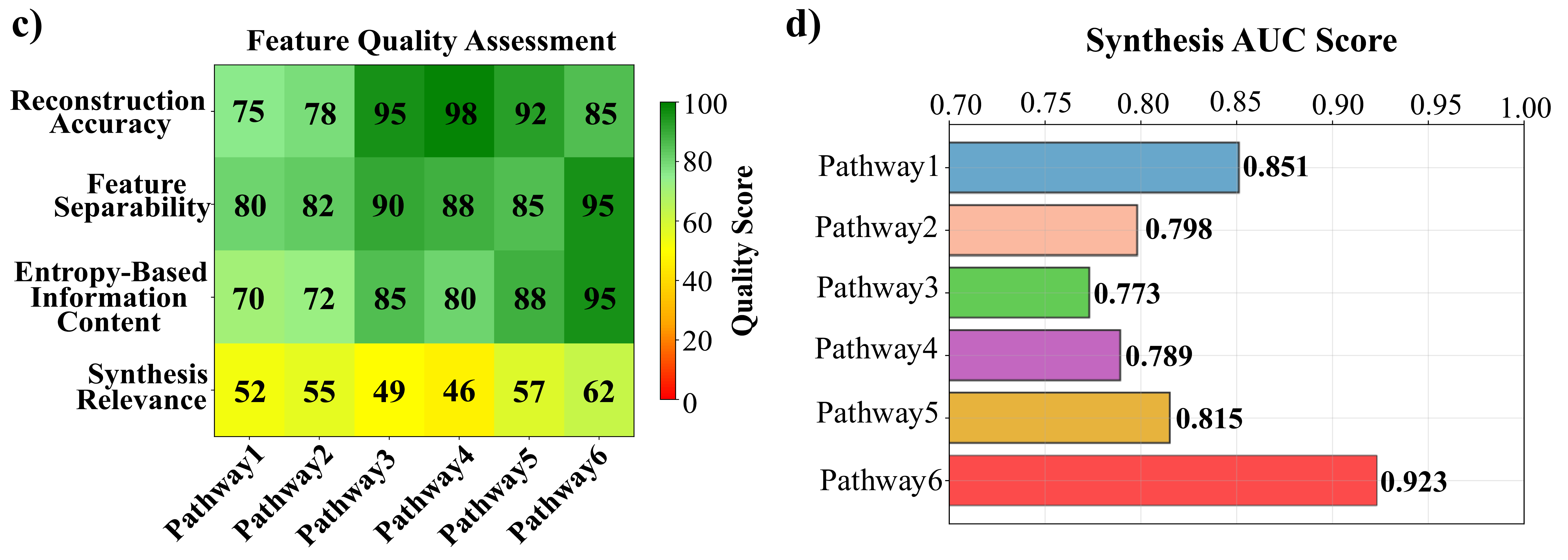}	
	\caption{Structure-aware feature extraction architecture for FTCP representation and performance evaluation. \textbf{a} Each crystal structure is encoded as a unified $399 \times 64$ FTCP tensor partitioned into six semantic blocks (elements, lattice parameters, atomic sites, site occupancy, reciprocal space, structure factors). \textbf{b} Six specialized neural network pathways for feature extraction, each optimized for its respective FTCP component. Pathways 3--6 employ transformer-inspired attention mechanisms (multi-head self-attention, cross-attention, graph attention), while Pathways 1--2 use linear transformations for their inputs. Each pathway produces a 256-dimensional feature vector (768 for Pathway 6), yielding a combined 2,048-dimensional representation. \textbf{c} Feature quality assessment heatmap displaying four evaluation metrics (reconstruction accuracy, feature separability, entropy-based information content, and synthesis relevance, each scored 0--100) across all six pathways, evaluated via self-supervised learning on the training set (N = 84,084). \textbf{d} Synthesis AUC scores for individual pathways, quantifying each pathway's discriminative power for synthesizability prediction on the training set. Pathway 6 (structure factors, multi-head attention + convolution) achieves the highest AUC (0.923).}
	\label{fig3}
\end{figure*}

One of the main elements of our approach is the representation of crystal structures through the Fourier-transformed crystal properties (FTCP), which encodes crystals in both real and reciprocal space. The FTCP partitions crystallographic information into six distinct components (elements, lattice parameters, atomic sites, site occupancy, reciprocal space features, and structure factors) as illustrated in Fig. \ref{fig3}a. Each component captures a complementary aspect of crystallography, ensuring that both atomic-level details and diffraction-like periodicity are preserved. Together, this systematic decomposition enables the representation of arbitrary crystal structures within a unified $399 \times 64$ tensor format, providing both compositional diversity and structural symmetry in a format naturally aligned with the physics of crystallography. FTCP construction is computationally lightweight. A systematic benchmark against standard atomic-graph representations shows that the median time for building an FTCP tensor from a CIF file is 2.045 ms per structure, compared to 1.770 ms and 1.899 ms for 5 Å all-neighbor and 8 Å k-nearest-neighbor graphs, respectively (Table S2, Figs. S9--S12). This near-equivalent cost, combined with the unique inclusion of reciprocal-space structure factors that are inaccessible to local graph convolutions, establishes FTCP as a practical and representationally rich alternative to atomic-graph inputs for crystal property prediction.

The inherent sparsity characteristics of the FTCP representation (as shown in Fig. S13) necessitate specialized feature extraction strategies to effectively capture the underlying patterns governing material synthesizability. As illustrated in Fig. \ref{fig3}b, the framework implements six specialized pathways: linear transformations for elemental composition and lattice parameters, attention mechanisms for atomic positions, cross- and graph-attention for occupancy states, convolutional and attention layers for reciprocal space, and convolution combined with multi-head attention for structure factors \cite{taniai2024crystalformer,du2024ctgnn,schmidt2021crystal,lee2025cast,ziletti2018insightful}. Each pathway architecture is optimized for its respective information content, ensuring that both local atomic relationships and global periodicity are learned. By employing advanced convolutional and attention-based layers aligned with the sparsity patterns of reciprocal space, the architecture efficiently compresses the FTCP input into a dense set of discriminative features (Fig. S14) suitable for downstream prediction.

Notably, transformer-inspired attention mechanisms constitute the core of this extraction process. Four of the six pathways (Pathways 3 through 6) employ multi-head self-attention, cross-attention, or graph attention, while only Pathways 1 and 2 rely on simpler linear transformations for their inputs. The attention-based pathways are also the dominant contributors to predictive performance, with Pathway 6 (multi-head attention combined with convolution) achieving the highest individual AUC of 0.923 (Fig. \ref{fig3}d).

The effectiveness of this multi-pathway approach is demonstrated through comprehensive feature quality assessment (Fig. \ref{fig3}c), where self-supervised learning was employed for all pathway-specific feature extraction to address the significant temporal distribution shift in synthesizability labels. Given imbalanced data and variation in synthesis rates across splits, supervised learning would likely bias features toward the historical synthesis patterns. By contrast, the self-supervised approach enables each pathway to learn robust crystallographic representations independent of synthesis labels, focusing on intrinsic structural and chemical patterns rather than temporal synthesis trends. The convergence behavior during this self-supervised training phase is highly uniform: all six feature-specific pathways demonstrate rapid initial convergence, reaching a low and stable loss value within approximately 40 epochs (Fig. S15). Moreover, the specialized neural network architectures achieve superior performance in capturing structural and chemical arrangements, with reconstruction accuracy scores ranging from 75\% to 98\% across different pathways. Feature separability metrics confirm that the extracted representations distinguish synthesizable from non-synthesizable structures, with most pathways scoring above 80\%, while entropy-based information content shows that the critical structural diversity is preserved despite dimensionality reduction. Importantly, the synthesis relevance scores highlight that reciprocal-space features (Pathways 5 and 6) contribute most directly to predicting synthesizability, whereas real-space features such as atomic sites and occupancies excel in accurate reconstruction. This complementarity reflects how the model balances faithful structural encoding with discriminative power, ensuring that both real- and reciprocal-space descriptors jointly support robust prediction.

The integration of pathway-specific features culminates in synthesis AUC scores (Fig. \ref{fig3}d) which quantitatively assess each pathway's contribution to the overall synthesizability prediction task. Pathway 6 achieves the highest performance (AUC = 0.923) confirming that structure factors provide the most discriminative information for distinguishing synthesizable materials. Elemental composition (Pathway 1, AUC = 0.851) and reciprocal-space features (Pathway 5, AUC = 0.815) also contribute strongly, whereas lattice, atomic site, and occupancy pathways show more modest AUC values ($\sim$0.77-0.80). This contrast highlights that reciprocal-space and composition-based descriptors are the primary drivers of predictive power, while real-space features primarily enhance reconstruction fidelity and structural realism. By combining these complementary strengths, the multi-stage feature extraction framework achieves robust and accurate synthesizability predictions across diverse material systems.

\subsection{Feature Selection}\label{subsec2_3}

The multi-pathway representation learning pipeline described above produced a combined feature space of 2,048 dimensions, with 256 latent descriptors extracted from each of five structural pathways and 768 from the reciprocal-space FTCP pathway as shown in Fig. S14. While this high-dimensional encoding captures rich information, it also risks redundancy and overfitting, particularly under the severe class imbalance of synthesizability data. Therefore, systematic dimensionality reduction becomes critical to extract the most discriminative features while maintaining model interpretability and computational efficiency. To identify the most informative subset of descriptors, we employed a feature selection stage based on Random Forest (RF) importance ranking \cite{breiman2001random}.

\begin{figure*}[!t]
	\centering
	\includegraphics[scale=1, height=0.60\textheight, width=1\textwidth, trim=9cm 17.5cm 3cm 2cm, clip]{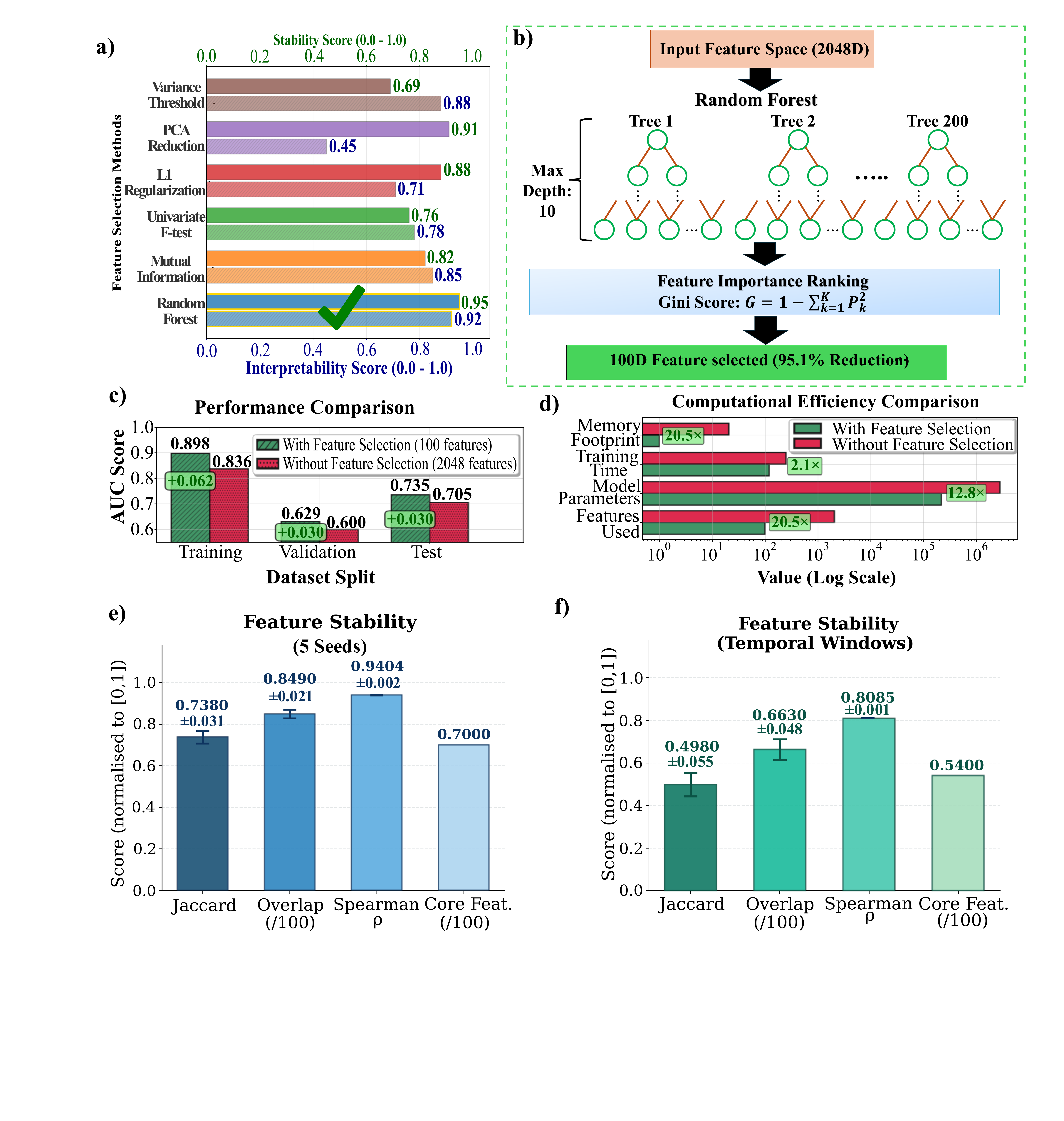}
	\caption{Random Forest-based feature selection and its impact on predictive performance, computational efficiency, and stability. \textbf{a} Comparison of six feature-selection methods using normalized stability and interpretability scores on a 0--1 scale, where higher values indicate better performance. Random Forest achieves the strongest overall balance. \textbf{b} Random Forest feature-selection pipeline. The 2,048-dimensional extracted feature space is ranked by Gini importance using an ensemble of 200 trees with maximum depth 10, and the top 100 features are retained, corresponding to a 95.1\% dimensionality reduction. \textbf{c} Predictive performance with and without feature selection across the training, validation, and test splits. AUC denotes area under the ROC curve. Feature selection improves training AUC from 0.836 to 0.898, validation AUC from 0.600 to 0.629, and test AUC from 0.705 to 0.735. \textbf{d} Computational-efficiency comparison on a logarithmic scale, showing reductions in features used, model parameters, training time, and memory footprint after reducing the representation. \textbf{e} Feature-selection stability across five independent random seeds normalized 0--1, summarized by Jaccard similarity, feature overlap, Spearman rank correlation of feature-importance scores, and the number of core features. Mean values are 0.738 \(\pm\) 0.031 (Jaccard), 0.849 \(\pm\) 0.021 shared features, 0.9404 \(\pm\) 0.002 (Spearman \(\rho\)), and 70 core features. \textbf{f} Feature-selection stability across temporal training windows containing 90\%, 95\%, and 100\% of the chronologically ordered training data. Mean values are 0.498 \(\pm\) 0.055 (Jaccard), 0.663 \(\pm\) 0.048 shared features, 0.809 \(\pm\) 0.001 (Spearman \(\rho\)), and 54 core features.}\label{fig4}
\end{figure*}

A comprehensive evaluation of feature selection methodologies was conducted to determine the optimal approach for this high-dimensional crystallographic feature space. As indicated in Fig. \ref{fig4}a, six distinct feature selection approaches were systematically assessed against two key criteria: stability, or whether the same features are consistently selected across data splits, and interpretability, or whether the selected features can be meaningfully related to crystallographic descriptors. Methods such as PCA, variance threshold, L1 regularization, univariate F-test, and mutual information demonstrated \cite{jolliffe2016principal,guyon2003introduction,tibshirani1996regression,zou2005regularization,kuhn2019feature,kraskov2004estimating,peng2005feature} trade-offs between the two metrics, often excelling in one but lagging in the other. In contrast, Random Forest-based feature selection achieved the best overall balance, with the highest stability score (0.95) and near-optimal interpretability (0.92). These advantages establish Random Forest as the recommended choice for extracting compact yet physically interpretable features in this application.

The ensemble nature of Random Forest provides robust feature importance estimates that are less susceptible to individual sample variations, while the Gini impurity-based importance scores offer direct interpretability regarding each feature's contribution to synthesizability classification \cite{hastie2009elements}. The selected configuration indicated in Fig. \ref{fig4}b utilized 200 trees with an optimized maximum depth of 10 (Fig. S16), achieving substantial dimensionality reduction by identifying 100 critical features from the original 2048-dimensional space, a 95.1\% reduction. The Gini impurity-based ranking mechanism ensures that retained features represent those most relevant for synthesizability prediction across diverse crystal structures. The benefits of this reduction are evident in both predictive performance and computational efficiency. Retaining only the 100 most informative features led to consistent improvements in AUC across all data splits, with gains of +0.062 in training, +0.030 in validation, and +0.030 in test sets (Fig. \ref{fig4}c). The effectiveness of this aggressive feature selection is demonstrated through comprehensive performance evaluation (Fig. \ref{fig4}c and Table S3). The reduced 100-feature representation maintains competitive predictive performance across all dataset splits, with training AUC scores of 0.898 (with selection) versus 0.836 (without selection), validation scores of 0.629 versus 0.600, and test scores of 0.735 versus 0.705. Notably, the reduced feature set improves generalization (smaller performance gaps between training and test sets), indicating lower overfitting. The computational advantages are equally substantial (Fig. \ref{fig4}d and Table S3): the streamlined feature set reduces memory usage by  $20.5 \times$, accelerates training by  $2.1 \times$, cuts the number of model parameters by $12.8 \times$, all while preserving discriminative capability. Together, these results demonstrate that Random Forest-based feature selection not only enhances prediction accuracy but also delivers the computational efficiency and interpretability needed for practical deployment in large-scale materials discovery workflows.

To ensure the robustness of the 100-dimensional feature set, we conducted two stability analyses (Fig. \ref{fig4}e,f). Across five independent random seeds with identical hyperparameters, the selected 100-feature sets exhibited high agreement, with an average Jaccard similarity of $0.738 \pm 0.031$, an overlap of $84.9 \pm 2.1$ features, and a Spearman rank correlation of $0.940 \pm 0.002$ on importance scores (Figs. S17--S18). A core set of 70 features appeared in every run, indicating that the majority of the selected descriptors are identified regardless of initialization. We further assessed stability across nearby temporal training windows (90\%, 95\%, and 100\% of the training data), which represents a more challenging perturbation because it alters the underlying data distribution. Under this condition, the Jaccard similarity was $0.498 \pm 0.055$ with 54 core features and a Spearman correlation of 0.809 (Figs. S19--S20), confirming that the overall importance landscape is preserved even when the training cutoff date shifts. Crucially, downstream performance remained insensitive to these variations. Test AUC across the five seeds was $0.732 \pm 0.010$ (Fig. S21), and across temporal windows it ranged from 0.717 to 0.736 (Fig. S22), demonstrating that the pipeline's predictions are robust to the specific feature set selected under different conditions.

\subsection{Synthesizability Prediction}\label{subsec2_4}

\begin{figure*}[!t]
	\centering
	\includegraphics[scale=1, height=0.50\textheight, width=1\textwidth, trim=0.5cm 23.7cm 15cm 0.2cm, clip]{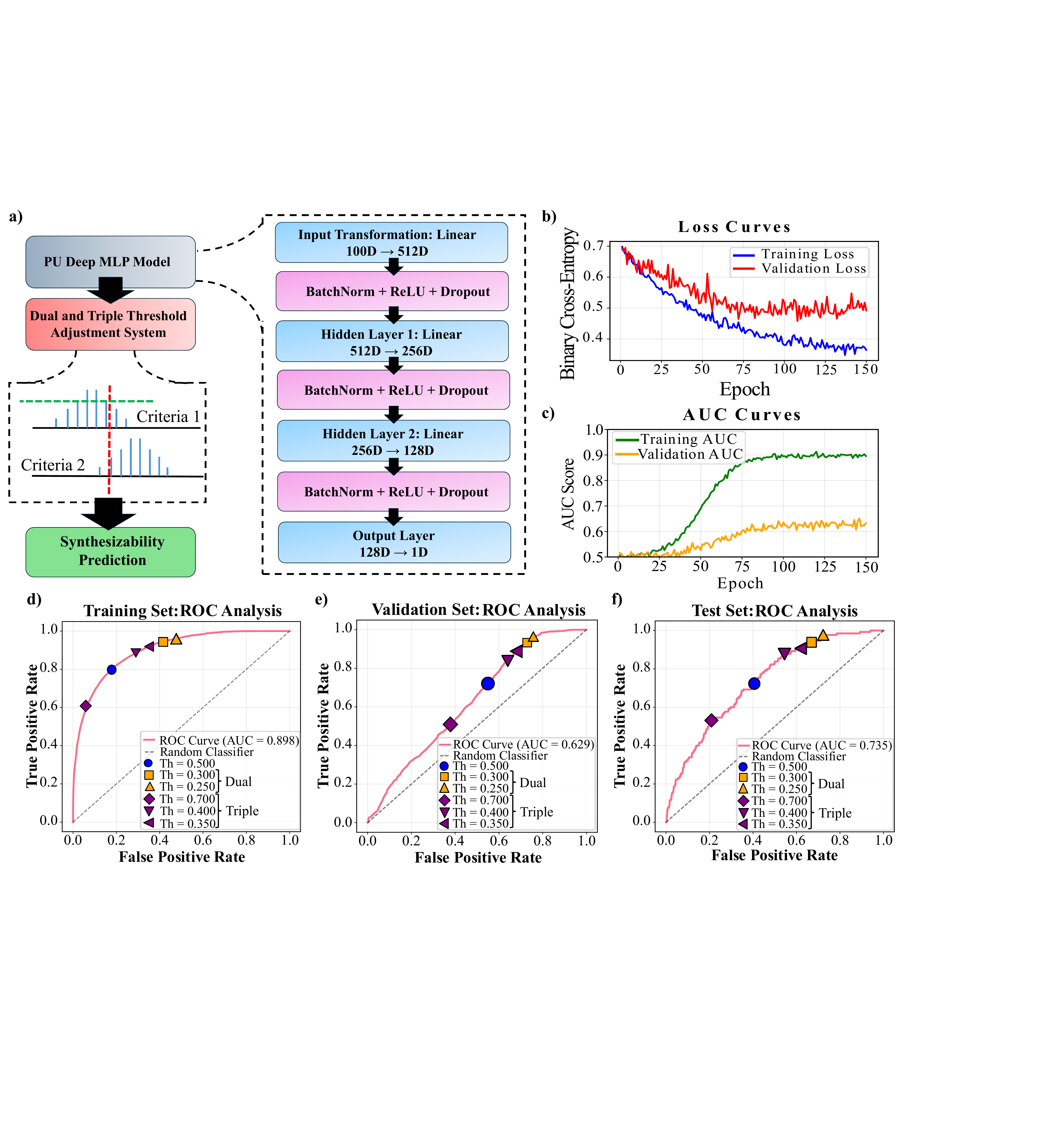}
	\caption{PU-MLP architecture, training dynamics, and thresholded ROC analysis for synthesizability prediction. \textbf{a} Deep multilayer perceptron (MLP) used for positive-unlabeled learning. The 100-dimensional selected feature vector is mapped through linear layers of size \(100 \rightarrow 512 \rightarrow 256 \rightarrow 128 \rightarrow 1\), with BatchNorm, ReLU, and Dropout applied after the hidden transformations, followed by dual- and triple-threshold adjustment systems for score-based decision making. \textbf{b} Binary cross-entropy loss over 150 training epochs for the training and validation sets, showing stable optimization and gradual convergence. \textbf{c} AUC evolution over 150 epochs for the training and validation sets. \textbf{d--f} ROC analysis for the training (\(N=84{,}084\)), validation (\(N=32{,}605\)), and test (\(N=12{,}784\)) partitions, respectively. Threshold markers indicate the standard cutoff (\(p=0.50\)), dual-threshold boundaries (\(p=0.30\) and \(0.25\)), and triple-threshold boundaries (\(p=0.70\), \(0.40\), and \(0.35\)). \label{fig5}
	}
\end{figure*}

With the optimized 100-dimensional feature set established through Random Forest selection, the reduced crystallographic descriptors serve as input to a deep multi-layer perceptron (MLP) architecture optimized for positive-unlabeled (PU) learning (Fig. \ref{fig5}a). The network employs a four-layer architecture with progressively decreasing dimensionality and incorporates batch normalization and dropout regularization to prevent overfitting in the severely imbalanced synthesizability dataset. Training convergence analysis demonstrates stable learning dynamics across 150 epochs (Fig. \ref{fig5}b,c). The binary cross-entropy loss exhibits consistent reduction for both training and validation sets, while AUC curves reveal rapid improvement in training performance. In synthesizability prediction, AUC is critical because it measures the model's ability to rank truly synthesizable compounds above the many unlabeled ones. A high AUC ensures viable candidates are prioritized, reducing the risk of overlooking promising materials. These outcomes reflect both the inherent difficulty of the task under extreme imbalance and the model's ability to retain generalizable signals across unseen structures. The PU learning framework was validated through two robust analyses. A sensitivity sweep over the class prior \(\pi_p\) shows that \(\pm 40\%\) perturbation produces only a 0.035 variation in test AUC (Figs.~S24--S26), and a controlled positive-to-unlabeled contamination experiment demonstrates that hiding up to 20\% of training positives reduces test AUC by only 0.022, with the model spontaneously recovering 58\% of hidden positives above the decision threshold (Figs.~S27--S30). Together, these results confirm that the pipeline's predictions are driven by learned structural representations rather than by the exact prior value or the completeness of training labels.

Receiver Operating Characteristic (ROC) analysis across all dataset splits reveals the model's discriminative capabilities under varying threshold configurations and highlights this pattern (Fig. \ref{fig5}d-f). Training set performance achieves an AUC of 0.898, demonstrating strong learning of synthesizability patterns from the historical data (2011--2018). The validation and test sets display shallower curves due to the scarcity of positives, which mirrors the increasing difficulty of discovering new compounds as chemical complexity rises. The validation set ROC analysis (Fig. \ref{fig5}e) shows compressed performance compared to training, with the optimal operating point shifting toward lower thresholds. This degradation reflects the drop from 49.8\% positive synthesis rate in training to only 7.9\% in validation, forcing the model to adapt to rare-event detection where viable compounds (positive cases) are hidden among a majority of unsynthesized (unlabeled) candidates. Importantly, the held-out test set (2019--2025) maintains an AUC of 0.735 despite the severe class imbalance (1.02\% positive), indicating robust predictive capability for future materials discovery. From a materials perspective, this robustness means that even under conditions where very few compounds are experimentally realized, the model can still elevate the most promising structures for synthesis trials. The trajectory of the ROC curve highlights that aggressive threshold reduction is essential in such scenario, ensuring that potentially synthesizable materials are not overlooked, a priority in experimental practice where missing a viable compound carries greater cost than testing additional false leads.

The implementation of adaptive threshold strategies directly addresses the severe class imbalance encountered in real-world synthesizability prediction. Standard binary classification at a 0.5 threshold proves inadequate for the extremely low base rate of successful synthesis in recent years. To overcome this, dual threshold adjustment (0.300 high, 0.250 low) was introduced (Fig. \ref{fig5}a), enabling explicit uncertainty quantification. This strategy achieved 97.6\% recall on the test set while designating 5.8\% of predictions as uncertain, reducing missed synthesizable materials from 28\% to 2.4\%. In materials discovery, this reduction is critical: false negatives correspond to overlooked opportunities for viable compounds that could otherwise be synthesized. These thresholds were calibrated on the validation set by optimizing a recall-weighted composite objective, ensuring that the selection reflects deployment priorities without exposure to test data. Sensitivity analysis across the full 0.25 to 0.40 operating range confirms that F1 remains stable between 61\% and 64\%, indicating that the model's overall predictive quality is robust to the exact threshold choice within this region. 

Building on this, multi-level thresholding transformed the binary classifier into a confidence-calibrated screening tool, enabling a graded prioritization of candidates according to synthesizability likelihood. The triple-threshold configuration provides four-confidence levels, maintaining 90.5\% recall while stratifying candidates into highly synthesizable, likely synthesizable, uncertain, and non-synthesizable groups. This risk-aware screening framework enables experimentalists to prioritize synthesis efforts based on available resources and tolerance for false positives. From a materials perspective, such calibration mirrors real laboratory practice, where promising but uncertain candidates may still merit exploration, while low-confidence predictions can be safely deprioritized. In this temporal PU setting, these confidence tiers should be interpreted operationally as score-based enrichment regions rather than as perfectly calibrated posterior probabilities, because the unlabeled pool may contain latent positives and the class prior shifts substantially between training and prospective deployment.

The ROC analysis confirms that these threshold strategies preserve the underlying model performance (AUC remains consistent) while enhancing practical deployment utility. By emphasizing high recall and introducing multi-level confidence tiers, the model functions as a decision-support tool for synthesis planning, ensuring that promising candidates are rarely overlooked while allowing flexible prioritization strategies aligned with different research goals. 

The temporal generalization from training data (49.8\% synthesis success rate) to test data (1.02\% success rate) represents a significant distributional challenge that the model addresses through robust feature learning and adaptive thresholding. Despite this drastic decline in success rates over time, the maintained AUC performance validates that self-supervised feature learning and PU classification can adapt to increasingly sparse discovery conditions. For materials scientists, this means the framework can still highlight synthesizable compounds even as research moves into more complex and less-explored chemical spaces.

\subsection{Relationship between Synthesizability and Thermodynamic Stability}\label{subsec2_5}
\begin{figure*}[!t]
	\centering
	
	\includegraphics[scale=1, height=0.41\textheight, width=1\textwidth, trim=0cm 15cm 0cm 0cm, clip]{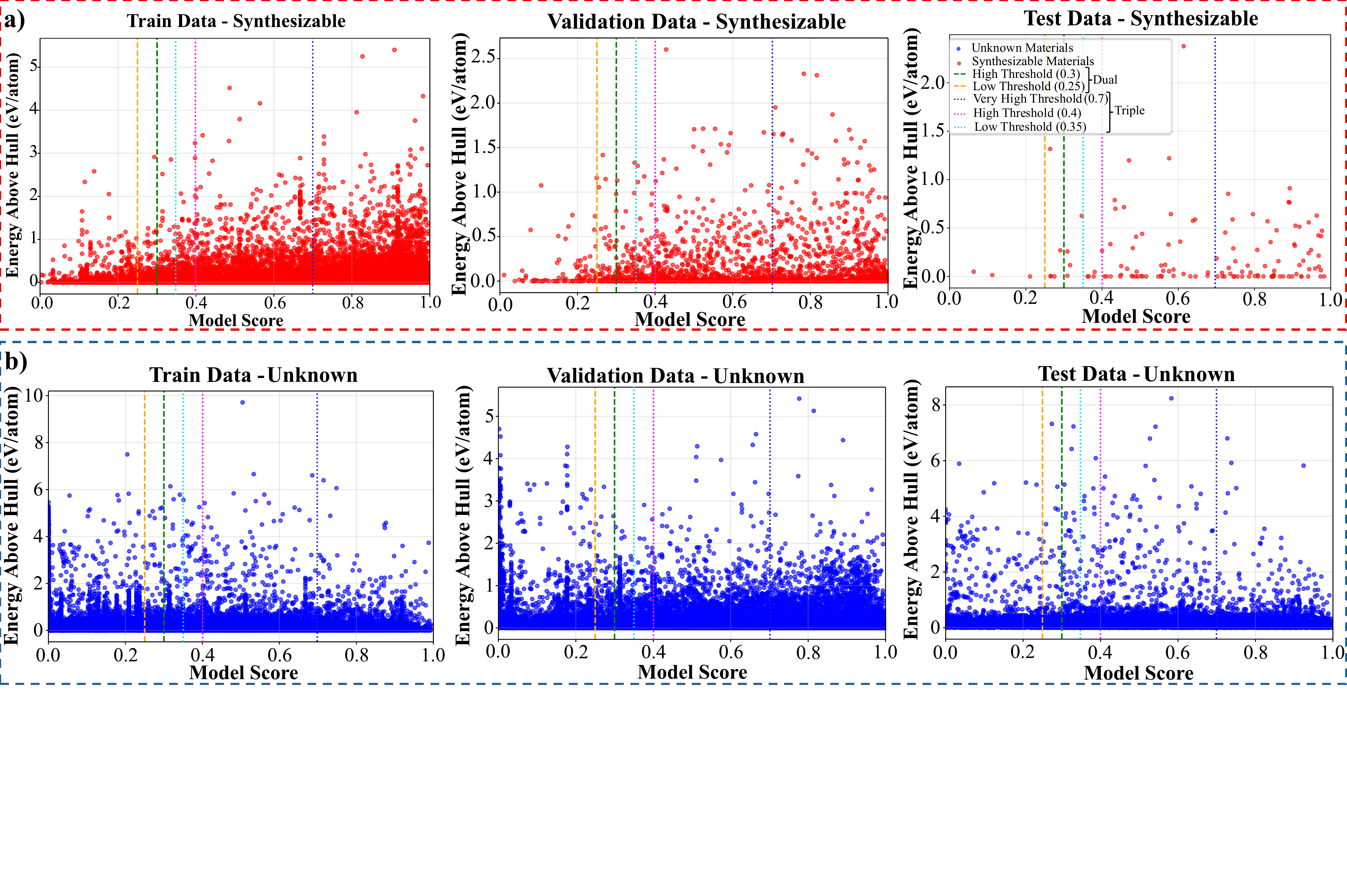}
	
	\vspace{0.01cm}
	
	\includegraphics[scale=1, height=0.22\textheight, width=1\textwidth, trim=0cm 38cm 0cm 0cm, clip]{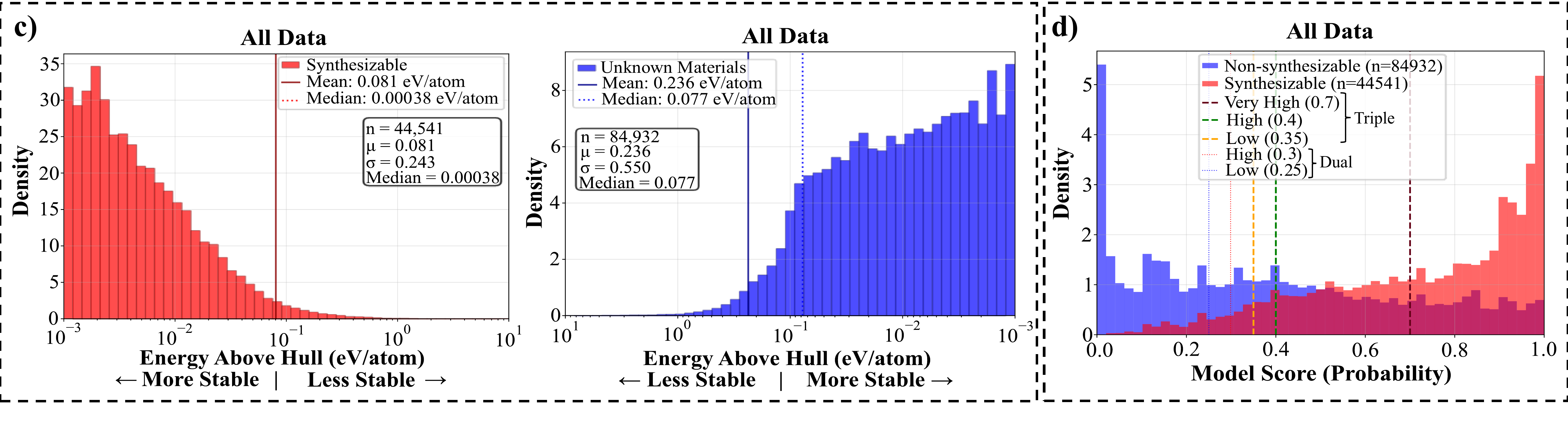}

	\caption{Relationship between synthesizability and thermodynamic stability across temporal splits and decision thresholds. \textbf{a} Energy above hull, \(E_{\mathrm{hull}}\), versus SyntheFormer score for experimentally synthesized materials in the training (\(N=41{,}849\)), validation (\(N=2{,}562\)), and test (\(N=130\)) partitions. Vertical dashed lines indicate the dual-threshold boundaries (\(0.25, 0.30\)) and triple-threshold boundaries (\(0.35, 0.40, 0.70\)). High-scoring synthesized materials are observed across a broad range of \(E_{\mathrm{hull}}\), including clearly metastable compounds, showing that synthesizability is not limited to near-hull phases. \textbf{b} Corresponding \(E_{\mathrm{hull}}\)-versus-score distributions for unlabeled materials in the training (\(N=42{,}235\)), validation (\(N=30{,}043\)), and test (\(N=12{,}654\)) partitions. Some unlabeled candidates occupy the same score--stability region as known synthesized materials, including high-scoring metastable regions, suggesting promising candidates for future experimental validation. \textbf{c} Distribution of \(E_{\mathrm{hull}}\) for all synthesized materials (\(N=44{,}541\)) and all unlabeled materials (\(N=84{,}932\)). Synthesized materials are concentrated closer to the convex hull (mean \(0.081\) eV/atom, median \(3.8\times10^{-4}\) eV/atom), whereas unlabeled materials are shifted to higher energies (mean \(0.236\) eV/atom, median \(0.077\) eV/atom), while still showing substantial overlap between the two populations. \textbf{d} Score distributions for synthesized and unlabeled materials over the full dataset, with dual and triple threshold markers superimposed. The score distribution is strongly bimodal, enabling uncertainty-aware screening through explicit score regions: low-confidence, intermediate/uncertain, and high-confidence regimes.}
	\label{fig6}
\end{figure*}

In a separate analysis spanning the complete materials dataset, Fig. \ref{fig6}a and Fig. \ref{fig6}b examine the relationship between SyntheFormer's predicted synthesizability scores and the thermodynamic descriptor energy above hull ($E_{\text{hull}}$). Analysis of confirmed synthesizable materials (Fig. \ref{fig6}a) reveals nuanced patterns across all dataset splits that extend beyond simple thermodynamic expectations. While the majority of synthesizable materials cluster in the low energy region (0-1 eV/atom above hull), notably some experimentally confirmed materials exhibit significantly higher energies above hull (up to 5+ eV/atom) that represent metastable phases or conditionally synthesized structures that the model successfully identifies despite their thermodynamic instability. This demonstrates that SyntheFormer captures synthesizability beyond what is implied by thermodynamic stability alone, a key advantage for identifying metastable yet experimentally accessible compounds.

The model demonstrates sophisticated discrimination by assigning varying confidence scores across the energy spectrum, which reflect the increased uncertainty in the materials' synthesizability. The threshold adjustment methods (dual and triple) effectively capture this complexity. On the test set, the dual-threshold approach captures 97.6\% of synthesizable compounds above the high threshold, while explicitly flagging 5.8\% of cases as uncertain. This ensures that nearly all viable candidates are prioritized, while leaving ambiguous cases for further evaluation. Extending this idea, the triple-threshold system introduces a finer-grained partition with four categories: highly synthesizable (above 0.70), likely synthesizable (0.40-0.70), uncertain (0.35-0.40), and non-synthesizable (below 0.35). This configuration maintains 90.5\% recall while distributing materials across multiple confidence levels, allowing experimentalists to adopt different strategies depending on their tolerance for risk and resource availability. In practical terms, the three-threshold calibration enables a risk-stratified screening pipeline, where high-confidence candidates can be fast-tracked for synthesis, while medium- or uncertain candidates remain accessible for exploratory work. The dual- and triple-threshold systems allow SyntheFormer to move beyond binary classification and into a multi-level confidence framework. Particularly striking is the model's performance on test data, where despite the severe class imbalance (1.02\% positive rate), it maintains robust discrimination of synthesizable materials across the full energy range.

At the same time, the unlabeled set (Fig. \ref{fig6}b) spans a much broader region, extending both to high energies and to low predicted probabilities. Despite containing numerous materials with favorable energies above hull (0-1 eV/atom), these remain unconfirmed experimentally. For these cases, SyntheFormer often assigns low or intermediate scores. This distinction demonstrates that synthesizability encompasses factors beyond thermodynamic stability, and the model has learned to recognize these subtle differences through its feature extraction from crystallographic descriptors rather than relying solely on energy-based metrics.

To contextualize SyntheFormer against a widely used proxy for attainability, we compared it with a DFT convex-hull threshold (synthesizable if $E_{\text{hull}}$ $<$ 0.1 eV/atom) as shown in Table S4. On the full dataset (129,473 structures), DFT correctly recovers 37,409 of 44,541 experimentally synthesized materials (recall = 84.0\% in total) but misses 7,132 synthesizable compounds. By contrast, SyntheFormer with dual thresholds correctly recovers 41,994 synthesized materials (94.3\% recall in total) and misses only 1,768, about $4 \times$ fewer false negatives than DFT at a comparable false-positive burden (48,054 vs 49,938). The triple-threshold setting strikes a different balance, correctly recovering 39,319 synthesized materials (88.3\% recall in total) and it reduces false positives by $\sim$30\% relative to DFT (38,377 vs 49,938) and missing 3,691, which is $\sim$$2 \times$ fewer false negatives than DFT (7,132 vs 3,691). Taken together, under our temporal split where the base rate of synthesis collapses to $\sim$1\% and the cost of false negatives is high, dual thresholds are most effective when the priority is minimizing missed opportunities, whereas triple thresholds are preferable when reducing experimental churn is paramount. These results substantiate the central claim that proximity to the convex hull is neither necessary nor sufficient for synthesizability, and that a structure-aware model can materially decrease the rate at which truly synthesizable compounds are overlooked.

Beyond accuracy, SyntheFormer provides uncertainty awareness, explicitly deferring borderline cases that DFT must classify. Dual thresholds flag 3.8\% of materials as uncertain (4,937/129,473), and triple thresholds 4.7\% (6,146/129,473), enabling expert triage where the base rate of success is low ($\sim$1\% in the test window).

Energy-based analysis of the complete dataset indicated in Fig. \ref{fig6}c and split data in Fig. S31--S32 reveals fundamental differences between material categories. Confirmed synthesizable materials exhibit a sharp peak near the convex hull with median equal to 0.00038 eV/atom that demonstrates the experimental bias toward thermodynamically stable phases. In contrast, unknown materials show a broader distribution (median: 0.077 eV/atom) extending to much higher energies that represent unexplored chemical space.

Stratified analysis reveals that SyntheFormer's primary added value over thermodynamic proxies lies in the metastable regime. Of the 7,132 experimentally synthesized materials with $E_{\text{hull}} \geq 0.1$ eV/atom that the DFT threshold misses by construction, SyntheFormer with dual thresholds recovers 6,762 (94.8\%), maintaining consistent recall of 90 to 96\% across all $E_{\text{hull}}$ regimes from mildly to strongly metastable (Figs. S33--S36). We further evaluated combined strategies in which an $E_{\text{hull}}$ prefilter removes high-energy materials before SyntheFormer scoring (Fig. S37). Tight prefiltering ($E_{\text{hull}} < 0.05$ eV/atom) improves precision to 61 to 64\% but reduces recall to 70 to 74\% by discarding metastable materials before the model can evaluate them. SyntheFormer alone achieves the highest recall (94.3\%) and captures all uniquely metastable discoveries, while a loose prefilter ($E_{\text{hull}} < 0.3$ eV/atom) preserves most of this advantage with modest precision gains. The optimal deployment strategy therefore depends on the experimental objective: maximum metastable discovery favors SyntheFormer alone, while precision-focused screening benefits from a combined approach.

The model score distribution analysis (Fig. \ref{fig6}d) shows the distribution of SyntheFormer model scores for all data, separated into synthesizable (red) and non-synthesizable (blue) classes. The bimodal distribution highlights that synthesizable compounds are strongly enriched at higher probabilities, while non-synthesizable materials dominate the lower score range. Under the dual-threshold configuration (0.25 low, 0.30 high), compounds above the high threshold are classified as synthesizable, those below the low threshold as non-synthesizable, and intermediate cases as uncertain. This calibration captures nearly all true positives while explicitly quantifying ambiguous cases. Instead of forcing binary decisions, SyntheFormer stratifies predictions into confidence tiers, enabling high-recall screening while explicitly quantifying uncertainty. By bridging data-driven predictions with stability descriptors, the framework delivers a principled guide for prioritizing new inorganic compounds for synthesis.

\begin{figure*}[!t]
	\centering
	\includegraphics[scale=1, height=0.36\textheight, width=1\textwidth, trim=2cm 30cm 0.1cm 21cm, clip]{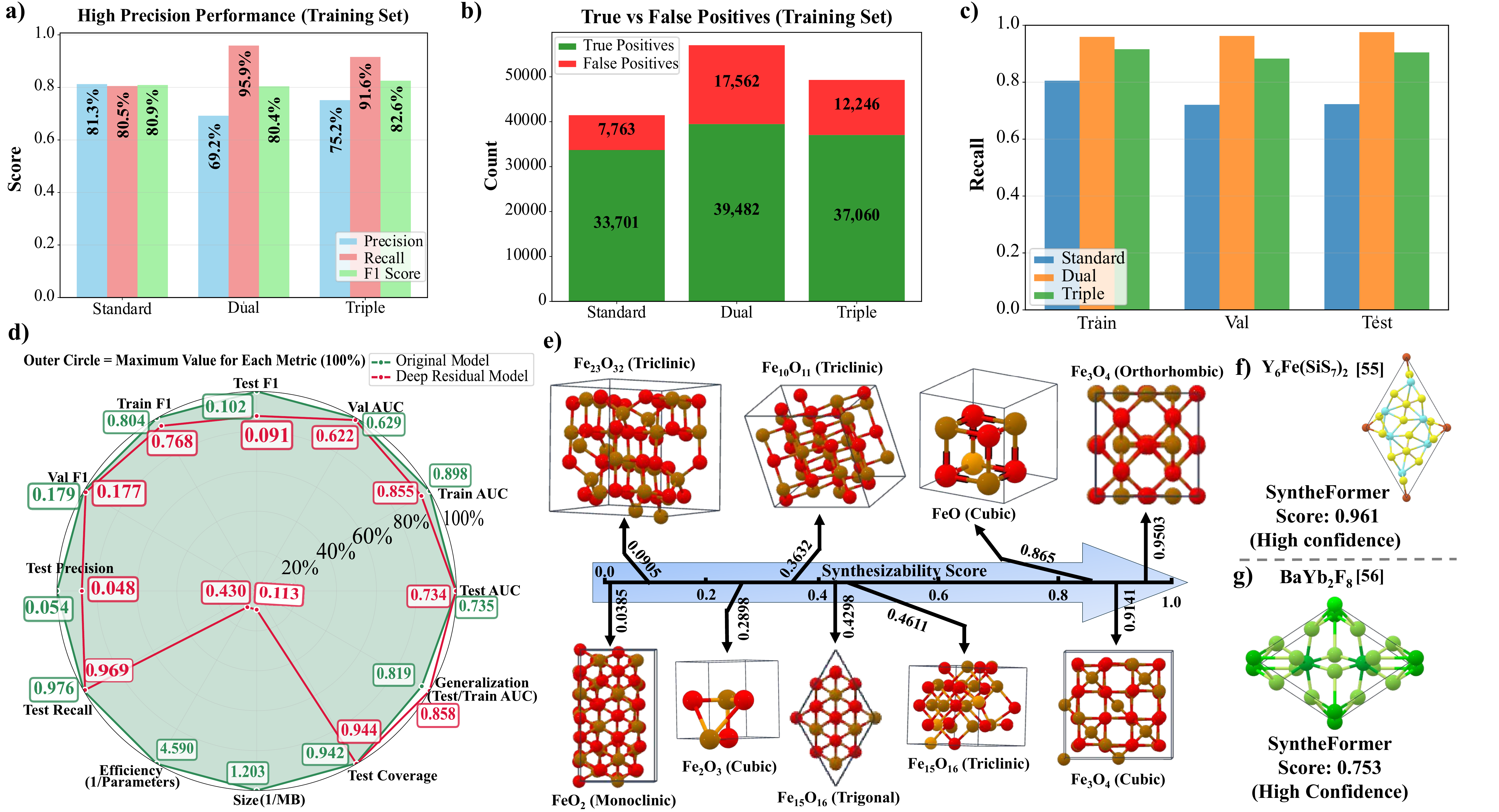}
	\caption{Threshold-strategy analysis, architecture comparison, and chemical interpretability of SyntheFormer. \textbf{a} Precision, recall, and F1 score on the training set for standard single-threshold, dual-threshold, and triple-threshold decision strategies. The dual-threshold strategy achieves the highest recall (95.9\%), whereas the triple-threshold strategy provides the strongest precision--recall balance with the highest F1. \textbf{b} Numbers of true positives and false positives on the training set for the same three strategies, showing that uncertainty-aware thresholding increases true-positive recovery while controlling false-positive burden. \textbf{c} Recall across the training, validation, and test splits for the three thresholding schemes. The dual-threshold strategy yields the highest recall on all three splits, while the triple-threshold strategy maintains a favorable balance between high recall and stricter prioritization.  \textbf{d} Radar-chart comparison between the original deep MLP and a deep residual-model baseline across multiple criteria. The original model achieves stronger overall performance while the residual baseline shows lower precision and lower overall screening utility. \textbf{e} Fe--O polymorph case study illustrating chemical interpretability of the model across multiple crystal systems and compositions. Structures are arranged according to their SyntheFormer synthesizability scores, showing that SyntheFormer distinguishes low-scoring and high-scoring Fe--O phases in a way that is chemically meaningful rather than determined solely by composition. \textbf{f,g} Crystal structures of two high-scoring materials that were unlabeled at the time of data curation but were subsequently synthesized experimentally in 2025: Y$_6$Fe(SiS$_7$)$_2$ (score 0.961) and BaYb$_2$F$_8$ (score 0.753). These prospective examples provide direct evidence that SyntheFormer can prioritize genuinely synthesizable materials ahead of experimental confirmation. \label{fig7}
	}
\end{figure*}

The comprehensive performance evaluation across threshold strategies reveals the transformative impact of adaptive classification on synthesizability prediction as shown in Fig. \ref{fig7}a,b). The base deep MLP model, trained with 100 selected features, demonstrates strong learning capacity with standard threshold classification achieving moderate performance (81.3\% precision, 80.5\% recall) on training data. However, performance degrades in temporally separated validation and test sets due to extreme class imbalance in recent years. Introducing dual-threshold calibration substantially enhances recall to 95.9\% while maintaining reasonable precision at 69.2\%, capturing 39,482 true positives compared to 33,701 under standard thresholding, albeit with increased false positives (17,562 versus 7,763). The triple threshold approach provides a balanced intermediate solution (75.2\% precision, 91.6\% recall), offering multiple confidence levels for risk-stratified screening. This strategic prioritization of synthesis opportunity capture over screening efficiency reflects the practical reality that missing a synthesizable material represents a greater cost than investigating additional candidates in materials discovery contexts, transforming a conventional classifier into a practical screening tool capable of operating under real-world deployment conditions.

The recall performance comparison (Fig. \ref{fig7}c) across all threshold calibration strategies confirms the consistent advantage of adaptive thresholding throughout the temporal distribution shift. Training recall improvements from 80.5\% (standard) to 95.9\% (dual) and 91.6\% (triple) demonstrate the method's effectiveness on balanced data. Crucially, these improvements persist across validation (72.1\% to 96.3\% to 88.3\%) and test sets (72.3\% to 97.6\% to 90.5\%), where the extreme class imbalance makes high recall critical for capturing rare synthesizable materials. The consistency of this improvement across different temporal periods validates the robustness of the threshold adaptation strategy for real-world deployment scenarios where synthesis success rates continue to decline as researchers explore increasingly challenging material targets.

In Fig. \ref{fig7}d, comparison with a deep residual architecture underscores the advantages of our streamlined MLP design for synthesizability prediction. Despite its simplicity and efficiency, the MLP consistently outperforms the residual variant across recall, AUC, and coverage metrics once dual-threshold calibration is applied. These gains are critical in a materials context, where high recall ensures that synthesizable compounds are rarely missed and broad coverage translates to a larger pool of viable candidates for laboratory screening. Training F1 scores (0.804 vs 0.768) demonstrate stronger pattern learning, while validation and test AUC scores (0.629 and 0.735, respectively) confirm generalization across the temporal distribution shift that reflect increasingly sparse synthesis rates.

Equally important, the MLP achieves this performance with only one-seventh the parameters of the residual baseline, reducing model size and improving computational efficiency. For materials screening workflows, this means that SyntheFormer can evaluate vast chemical spaces rapidly without requiring excessive computational resources, a practical necessity for integration into high-throughput discovery pipelines. The residual model's lower stability further demonstrates that increasing architectural depth does not necessarily capture more meaningful crystallographic features. Instead, the multi-pathway FTCP representation and targeted feature selection provide a more physically grounded encoding of crystallographic descriptors, eliminating the need for deeper architectures while maintaining predictive robustness. This alignment between model efficiency, interpretability, and predictive power makes the streamlined MLP a more suitable choice for real-world deployment in materials discovery efforts.

The iron oxide polymorphs case study (Fig. \ref{fig7}e) demonstrates the model's chemical interpretability and structural discrimination capabilities. This discrimination reflects genuine differences in synthetic accessibility where common iron oxides like halite ($\mathrm{Fe O}$ cubic) and magnetite ($\mathrm{Fe_3 O_4}$ orthorhombic) receive appropriately high scores, while less stable or more complex structures show graduated confidence levels. The diversity of predicted structures across different crystal systems (monoclinic, cubic, trigonal, orthorhombic) validates that the model has learned systematic crystallographic principles rather than memorizing specific compositions. Together, these results establish SyntheFormer's MLP core, enhanced by multi-threshold calibration, as a reliable predictor of synthesizability across chemical spaces and temporal regimes. The framework maintains high discriminative power, minimizes missed opportunities, and enables practical triaging of candidate materials for laboratory synthesis.

As a direct prospective validation, we cross-referenced high-scoring unlabeled predictions against the most recent experimental literature published well after our primary data curation. We highlight two significant discoveries that validate the predictive power of SyntheFormer. First, a quaternary rare-earth chalcogenide Y$_6$FeSi$_2$S$_{14}$ was assigned a synthesizability probability of 0.961 by our model (Fig. \ref{fig7}f). This material, which possesses a complex hexagonal structure with specialized magnetic sublattices, was experimentally realized in 2025 (Gamage et al. \cite{gamage2025interplay}), confirming its viability. Second, BaYb$_2$F$_8$, a material investigated for fluoride-ion conductivity, received a high-confidence score of 0.753 and was subsequently synthesized by Matsui et al. \cite{matsui2025tailoring} in 2025 (Fig. \ref{fig7}g). Both scores exceed the high-confidence threshold, meaning these materials would have been flagged as highly synthesizable by SyntheFormer before their experimental realization, providing direct evidence that the model captures generalizable structural signatures of synthesizability beyond the training data.

\subsection{Comparative Performance Evaluation}\label{subsec2_6}
To rigorously position SyntheFormer within the existing landscape of synthesizability prediction, we benchmarked it against representative alternative models evaluated under the same temporal split used throughout this work (Table~\ref{tab1}). All models were trained on identical data partitions and evaluated on the held-out test set comprising 12,784 materials with only 130 positives (1.02\% positive rate). For fair comparison, recall, precision, and F1 are reported at each model's best-F1 operating point calibrated on the validation set and held fixed during test evaluation. We additionally report AUPRC, a threshold-independent metric that is more informative than AUC under severe class imbalance because it directly measures a model's ability to enrich true positives among its highest-ranked predictions. The random AUPRC baseline for a 1.02\% positive rate is 0.010.

Among existing synthesizability prediction models, we re-implemented and trained SynCoTrainMP \cite{amariamir2025syncotrain}, a SchNet-based message-passing GNN with three interaction blocks; Synthesizability-PU-CGCNN \cite{jang2020structure}, a 100-bag bootstrap ensemble of Crystal Graph Convolutional Neural Networks where the final score is the mean probability across all bags; and XIE-SPP \cite{davariashtiyani2021predicting}, a 3D-CNN operating on \(128^3\) voxel images encoding atomic number, group, and period within a 50~\AA{} bounding box. SyntheFormer achieves the highest test AUC (0.735) and test AUPRC (0.099), with the latter being approximately 9.9 times the random baseline and nearly double the next-best synthesizability model (XIE-SPP, AUPRC = 0.053). At the best-F1 operating point, SyntheFormer and XIE-SPP recover the same number of positives (121 of 130, recall = 0.931), but SyntheFormer's precision is 7-fold higher (0.103 versus 0.015), yielding an F1 of 0.186 compared to 0.029. Cross-split analysis (Tables S5--S6 and Figs. S38--S39) reveals that SyntheFormer maintains a consistent performance trajectory across temporal partitions (training AUC 0.898, validation 0.629, test 0.735), whereas competing models show either overfitting (XIE-SPP: training 0.870, validation 0.540) or poor temporal generalization (SynCoTrainMP: validation 0.796, test 0.697). These patterns confirm that the self-supervised feature extraction pipeline, combined with Random Forest selection and PU-aware classification, provides robust generalization under the severe temporal distribution shift that characterizes real-world synthesizability prediction.

To quantify the benefit of structure-aware features, we additionally trained three composition-only baselines that use no crystal structure information and operate only on the chemical formula (Table 1). All three employ the same MLP architecture and PU learning loss as SyntheFormer, ensuring that the input representation is the only variable. B1 encodes raw integer element counts (dimension 87), B2 encodes normalized element fractions (dimension 87), and B3 uses Meredig-style descriptors augmenting atomic fractions with 32 elemental property statistics \cite{ward2018matminer} (dimension 135). Despite spanning a range from minimal to rich composition features, all three baselines converge to a narrow test AUPRC range of 0.03 to 0.037, compared to SyntheFormer's 0.099. Even B3, the most informative composition descriptor, achieves an AUPRC that is 3.6 times the random baseline and 2.7 times lower than SyntheFormer. At the best-F1 operating point, SyntheFormer achieves 10.3\% precision compared to 2.8\% for the best composition model (B1). Cross-split analysis (Tables S5--S6 and Figs. S40--S41) further reveals that all composition baselines achieve validation AUC at or below 0.58, compared to SyntheFormer's 0.629, indicating that the geometric and reciprocal-space information encoded in FTCP captures physical constraints that generalize across temporal windows in a way that composition features alone cannot. 

Finally, we benchmarked SyntheFormer against two widely utilized structure-aware Graph Neural Networks: MEGNet and CGCNN. MEGNet \cite{chen2019graph}, a message-passing GNN with three interaction blocks and Set2Set readout (169,937 parameters), and CGCNN \cite{xie2018crystal}, with three graph-convolutional layers and 92-dimensional atom embeddings (80,962 parameters), were both trained from scratch on our temporal split using the same nnPU loss function and early-stopping protocols as our primary framework. As summarized in Table 1, these GNNs exhibit limited generalizability in the context of synthesizability discovery. MEGNet and CGCNN achieve test AUPRC values of 0.0309 and 0.0317, respectively, falling below the 0.099 achieved by SyntheFormer. In practical terms, SyntheFormer provides an $9.9\times$ improvement over the random baseline, whereas the standard structural models provide approximately 3.1-fold improvement. Furthermore, at the optimal operating points calibrated on validation data, SyntheFormer maintains a precision of 0.1031, whereas MEGNet and CGCNN reach only 0.0130 and 0.0179. Validation performance further reveals that both GNNs operate near random level (AUC 0.538 and 0.513, respectively), compared to SyntheFormer's 0.629 (Tables S5--S6, Figs. S42--S43). Since all models were trained on the same data using structural inputs, the performance discrepancy suggests that the local coordination environments captured by standard graph message-passing are insufficient for modeling the complex, long-range periodicities inherent in synthesis. The performance gap therefore confirms that the marginal value of SyntheFormer lies not in using structural information alone, but in the combination of FTCP's reciprocal-space encoding with self-supervised multi-pathway feature extraction, which together produce representations that generalize robustly under temporal distribution shift.

\begin{table*}[t]
	\centering
	{\let\raggedright\relax
	\caption{Comparative performance of SyntheFormer against representative synthesizability prediction models, composition-only baselines, and standard GNN baselines under the same temporal split, evaluated on the prospective test set (\(N=12{,}784\), including 130 positives; 1.02\% positive rate). Recall, precision, and F1 are reported at each model's best-F1 threshold selected on the validation set and then held fixed for test evaluation. SyntheFormer achieves the highest test AUC (0.7350) and AUPRC (0.099), corresponding to nearly a ten-fold enrichment over the random AUPRC baseline. At the best-F1 operating point, SyntheFormer and XIE-SPP achieve the same recall (0.9308), but SyntheFormer attains substantially higher precision (0.1031 vs 0.0147) and a much higher F1 (0.1860 vs 0.0289).}}
	\label{tab1}
	\begin{threeparttable}
	\footnotesize
	\setlength{\tabcolsep}{4pt}
	\begin{tabular}{p{2.0cm}lccccc}
	\toprule
	\textbf{Category} & \textbf{Model} & \textbf{AUC} & \textbf{AUPRC} & \textbf{Recall} & \textbf{Precision} & \textbf{F1} \\
	\midrule
	\multirow{4}{2.0cm}{\centering\textit{Structure-\\aware models}}
	& \textbf{SyntheFormer (ours)}   & \textbf{0.7350} & \textbf{0.099}  & 0.9308 & \textbf{0.1031} & \textbf{0.1860} \\
	& SynCoTrainMP (SchNet)\cite{amariamir2025syncotrain} & 0.6971 & 0.0259 & 0.2308 & 0.0294 & 0.0521 \\
	& PU-CGCNN\cite{jang2020structure}              & 0.7253 & 0.0358 & 0.4923 & 0.0206 & 0.0396 \\
	& XIE-SPP\cite{davariashtiyani2021predicting}               & 0.7216 & 0.053  & 0.9308 & 0.0147 & 0.0289 \\
	\midrule
	\multirow{3}{2.0cm}{\centering\textit{Composition-\\only baselines}}
	& MLP + ElemCount       & 0.6947 & 0.0359 & 0.0923 & 0.0280 & 0.0429 \\
	& MLP + ElemFraction    & 0.7128 & 0.0303 & 0.7462 & 0.0155 & 0.0304 \\
	& MLP + Meredig         & 0.6896 & 0.0366 & 0.8077 & 0.0121 & 0.0239 \\
	\midrule
	\multirow{2}{2.0cm}{\centering\textit{Standard GNN\\baselines}}
	& MEGNet\cite{chen2019graph}                & 0.7002 & 0.0309 & \textbf{0.9385} & 0.0130 & 0.0256 \\
	& CGCNN\cite{xie2018crystal}                 & 0.7259 & 0.0317 & 0.7308 & 0.0179 & 0.0350 \\
	\bottomrule
	\end{tabular}
	\end{threeparttable}
	\end{table*}

As established in the preceding benchmarks, SyntheFormer achieves a Test AUPRC of 0.099, outperforming other baselines which range from 0.025 to 0.053. This nearly 10-fold enrichment over the random baseline (0.0102) is visually confirmed by the Precision-Recall curve provided in Fig. S44. The analysis reveals a precision plateau where the model maintains a 10.3\% to 10.8\% precision for recall values up to 93.1\%. This stability is followed by a sharp decline in precision, providing a clear quantitative boundary for our high-confidence discovery thresholds and confirming that the model’s ranking ability is robust even when the positive class is exceptionally rare.

\section{Discussion}\label{sec3}
The development of SyntheFormer represents a paradigm shift in computational materials discovery by directly addressing the fundamental challenge of synthesizability prediction through a data-driven approach. By reformulating materials discovery as a positive-unlabeled learning problem and leveraging the entire spectrum of experimentally realized crystalline materials, our framework achieves unprecedented accuracy in identifying synthetically accessible compounds while operating under the severe class imbalance that characterizes real-world materials exploration.

The superior performance of SyntheFormer over traditional approaches stems from its ability to learn complex, non-linear relationships between crystallographic features and synthetic accessibility that extend far beyond simple thermodynamic or charge-balancing criteria. While formation energy calculations capture only thermodynamic favorability and charge-balancing approaches rely on rigid oxidation state constraints, our structure-aware FTCP feature extraction combined with self-supervised learning enables the model to discover subtle patterns governing synthesizability across diverse chemical systems. This is particularly evident in the model's ability to correctly identify metastable phases with high energy above hull values that have been experimentally synthesized, demonstrating that synthesizability encompasses factors beyond thermodynamic stability alone. In a head-to-head baseline (Table S4), a DFT hull proxy misses 7,132 ICSD-confirmed materials, whereas SyntheFormer with dual thresholds misses 1,768 that is about four-fold fewer false negatives at a comparable false-positive burden. Comprehensive benchmarking against published synthesizability models under the same temporal split confirms this advantage quantitatively, with SyntheFormer achieving the highest test AUPRC and nearly doubling the next-best method while training in a fraction of the computational time. Stratified analysis confirms that this advantage concentrates in the metastable regime, where SyntheFormer recovers 94.8\% of the 7,132 synthesized materials with $E_{\text{hull}} \geq 0.1$ eV/atom that the DFT criterion misses entirely (Figs. S33--S36).

The temporal generalization capabilities of SyntheFormer address a critical limitation in materials discovery, the ability to predict the synthesizability of genuinely novel compounds that differ significantly from historical training data. The maintained AUC performance of 0.735 on test data from 2019--2025 despite the dramatic shift from 49.8\% to 1.02\% positive synthesis rates validates the robustness of our self-supervised feature learning approach. This temporal stability is crucial for practical deployment, where materials discovery efforts increasingly target complex compositions with lower inherent synthesis probabilities. Under a forward temporal split where the positive base rate collapses to $\sim$1\%, dual thresholds sustain high recall (94.3\% over all positives), while the triple-threshold mode reduces false positives by $\sim$30\% relative to the DFT proxy. An inherent concern in temporal PU settings is that the unlabeled training pool contains structures that will be synthesized after the training cutoff. A controlled contamination experiment in which up to 20\% of training positives were deliberately relabeled as unlabeled shows that test AUC drops by only 0.022, and the model spontaneously recovers the majority of hidden positives with high predicted probability (Figs. S27--S30), confirming that SyntheFormer learns generalizable structural features of synthesizability rather than memorizing specific labeled instances. Direct prospective validation reinforces this conclusion: two materials that were unlabeled in our dataset, Y$_6$Fe(SiS$_7$)$_2$ and BaYb$_2$F$_8$, have since been experimentally synthesized and received SyntheFormer scores of 0.961 and 0.753, respectively, both well above the high-confidence threshold.

The adaptive threshold strategies represent a key innovation that transforms standard binary classification into a practical materials screening tool. The dual and triple threshold configurations address the asymmetric costs inherent in materials discovery, where missing a synthesizable material (false negative) incurs far greater opportunity costs than investigating an ultimately unsuccessful candidate (false positive). By achieving 97.6\% recall on test data while providing explicit uncertainty quantification, these threshold strategies enable materials scientists to operate with confidence levels appropriate to their resource constraints and risk tolerance. Unlike fixed DFT cutoffs, SyntheFormer can explicitly defer borderline cases (3.8-4.7\% of entries) for expert review, providing calibrated uncertainty rather than forced binary decisions.

While end-to-end architectures are popular in deep learning, the extreme sparsity of the $399\times64$ FTCP tensor and the severe class imbalance of the positive-unlabeled dataset necessitated a modular approach. By separating representation learning (the multi-pathways framework) from feature selection (Random Forest) and classification (MLP), we achieved a framework that is both computationally efficient and robust to temporal distribution shifts. This modular design also enables independent evaluation and interpretability at each stage, including pathway-specific metrics and Gini-based feature importance ranking, providing insight into which crystallographic aspects drive the predictions in a way that would be difficult to obtain from a fully end-to-end architecture.

The interpretability analysis reveals that SyntheFormer has learned fundamental chemical principles without explicit instruction. The model's differential application of charge-balancing constraints to ionic versus covalent compounds, its recognition of periodic table relationships through learned embeddings, and its ability to exploit chemical analogy all demonstrate that the framework captures genuine chemical knowledge rather than spurious statistical correlations. This emergent chemical understanding provides confidence that the model's predictions are grounded in physically meaningful principles.

The integration of SyntheFormer into materials discovery workflows offers transformative potential for accelerating the identification of novel functional materials. The framework's computational efficiency extends to every stage of the pipeline, from FTCP construction, which incurs negligible overhead compared to atomic-graph alternatives, to downstream inference requiring only milliseconds per prediction compared to hours or days for DFT calculations, enabling screening of vast chemical spaces previously inaccessible to systematic exploration. When incorporated into inverse design pipelines, SyntheFormer ensures that computationally generated candidates are synthetically realistic with reliable confidence rates, improving the success rate of subsequent experimental validation efforts. Because SyntheFormer provides score-stratified confidence estimates and uncertainty flags, it is straightforward to tune operating points to laboratory priorities-maximizing recall when the goal is discovery breadth, or tightening specificity (via triple thresholds) to curb experimental churn.

The performance advantages demonstrated by SyntheFormer extend beyond computational approaches to human expert judgment. The model's consistent accuracy across diverse chemical families, highlights the value of training on the entire corpus of synthesized materials rather than relying on limited domain-specific heuristics. By capturing the full diversity of prior synthesis outcomes, SyntheFormer provides a data-driven perspective that systematically corrects for human cognitive biases and incomplete theoretical models. This capacity to generalize across unexplored compositional and structural regimes positions the framework as a powerful complement to both expert intuition and physics-based simulations.

Taken together, these advances establish SyntheFormer as more than a predictive model---it represents a methodological shift in how synthesizability can be quantified, interpreted, and operationalized in computational pipelines. Looking forward, the continued growth of materials databases and the development of new synthetic methodologies will further enhance SyntheFormer's predictive capabilities. The framework's data-driven foundation allows it to naturally incorporate new synthetic knowledge as it becomes available. The modular architecture of feature extraction and adaptive thresholding provides a foundation for extensions to other material classes and properties. The comparative evaluation provides direct empirical support for this design, as standard end-to-end GNN architectures (MEGNet and CGCNN) trained with the same PU loss on the same structural data achieve test AUPRC more than 3 times lower than SyntheFormer's modular pipeline.

The broader implications of this work extend beyond synthesizability prediction to demonstrate the power of positive-unlabeled learning approaches for materials science applications where negative examples are scarce or poorly defined. Many important materials properties such as stability under operating conditions, processability, or scalability share the characteristic that positive examples are well-documented while negative evidence is sparse. The methodological framework developed here provides a template for addressing these challenges across diverse materials discovery applications.

SyntheFormer establishes synthesizability prediction as a mature complement to traditional computational materials discovery methods. The combination of temporal robustness, interpretability, uncertainty quantification, and practical thresholding transforms synthesizability prediction from a theoretical curiosity into a deployable tool for accelerating experimental progress, ensuring these efforts are focused on the most promising candidates within the vast landscape of theoretically possible materials. By bridging the gap between computational prediction and experimental realization, SyntheFormer offers a foundation for a new era of guided discovery in chemistry and materials science. Moreover, under a stringent, forward-looking temporal split, SyntheFormer reduces missed synthesized materials by roughly four-fold relative to a DFT hull proxy while enabling ~30\% fewer false positives at a more conservative operating point (Table S3), establishing a practical, confidence-aware standard for screening at scale.

\section{Method}\label{sec4}
\subsection*{Dataset construction and temporal splitting}
The synthesizability prediction framework was developed using materials data from the Materials Project database combined with temporal splitting to simulate realistic deployment conditions. The dataset comprises 129,473 inorganic crystalline materials spanning binary, ternary, and quaternary compositions from 2011--2025. Positive labels correspond to entries cross-referenced with the Inorganic Crystal Structure Database (ICSD), ensuring only experimentally synthesized compounds are marked as positive examples, while remaining entries are treated as unlabeled in the positive-unlabeled (PU) learning framework.

Materials were temporally partitioned with training data from 2011--August 2018 (84,084 entries, 49.8\% positive), validation data from September--December 2018 (32,605 entries, 7.9\% positive), and test data from 2019--2025 (12,784 entries, 1.02\% positive). This temporal splitting prevents data leakage and reflects the increasing difficulty of materials synthesis over time. The partition is applied before any processing, and every learned transformation in the pipeline (FTCP pathway training, Random Forest feature selection, StandardScaler normalization, MLP weights, and threshold calibration) is derived exclusively from training-period data, with the test set serving as a sealed prospective holdout evaluated once after all decisions are frozen.

\subsection*{FTCP representation}
Crystal structures were encoded using Fourier-Transformed Crystal Properties (FTCP) representation, partitioning crystallographic information into six components: elemental composition (indices 0:102, 0:4), lattice parameters (103:104, 0:4), atomic sites (105:204, 0:3), site occupancy (205:304, 0:4), reciprocal space features (305, 4:63), and structure factors (306:398, 0:63). This yields unified $399\times64$ tensor representations capturing both real and reciprocal space information. To quantify the computational cost of FTCP construction relative to other representations, we benchmarked FTCP against two atomic-graph methods (a 5 Å all-neighbor graph and an 8 Å k-nearest-neighbor graph) on 1,000 randomly sampled Materials Project structures. Each structure was processed 15 times per method to obtain robust timing statistics. 

\subsection*{Structure-aware feature extraction}
Due to the sparsity of FTCP, a multi-pathway feature extraction pipeline was implemented. The structure-aware feature extraction architecture employed six specialized neural network pathways, each optimized for specific FTCP components. Self-supervised learning was applied to address temporal distribution shifts in synthesis labels, enabling pathway-specific networks to learn crystallographic representations independent of synthesis outcomes. Architecture details include multi-head attention mechanisms for spatial relationships, graph neural networks for occupancy analysis, and convolutional layers for reciprocal space processing. The five structural pathways produced 256 features each, while the reciprocal-space pathway produced 768 features, yielding a combined 2048-dimensional feature space.

\subsection*{Feature selection}
Random Forest-based feature selection reduced the 2048-dimensional combined feature space to 100 dimensions. The selector employed 200 trees with maximum depth 10, achieving 95.1\% dimensionality reduction. This reduction substantially improves generalization and computational efficiency while retaining discriminative power. Gini impurity-based importance ranking identified features with highest contribution to synthesizability classification:
\begin{equation}
	G=1-\sum_{k=1}^{K} P_{k}^{2}
\end{equation}

where $G$ represents Gini impurity, $K$ is the number of classes, and $P_{k}$ is the proportion of samples belonging to class $k$.

Feature selection stability was assessed along two axes. First, five independent runs with different random seeds were performed using the same RF hyperparameters on 20,000 training samples, and the resulting feature sets were compared via Jaccard similarity, pairwise overlap counts, and Spearman rank correlation of importance scores. Second, temporal stability was evaluated by varying the training window to 90\%, 95\%, and 100\% of the training data and comparing the selected feature sets across these conditions.

\subsection*{Model architecture and training}
The synthesizability predictor is a four-layer multi-layer perceptron (MLP) optimized for PU learning. The architecture was:
\begin{equation}
	100\rightarrow 512\rightarrow 256\rightarrow 128\rightarrow 1,
\end{equation}
with ReLU activations, batch normalization, and dropout (0.2, 0.2, 0.1). Training followed a risk estimation approach for PU learning. Let $\mathcal{P}$ be the positive set, $\mathcal{U}$ the unlabeled set, and $\pi_p$ the class prior. The loss function is:
\begin{equation}
	\zeta_{PU}=\pi_{p}\sum_{x\sim p}^{}\left [ \ell(f(x),1) \right ]+max(0,\sum_{x\sim u}^{}\left [ \ell(f(x),0) \right ])-\pi_{p}\sum_{x\sim p}^{}\left [ \ell(f(x),0) \right ]
\end{equation}

Where \(\ell\) is the binary cross-entropy loss and \(\pi_p = 0.50\) was estimated from the empirical positive rate of the temporal training partition (49.8\%). A sensitivity sweep across \(\pi_p \in \{0.30, 0.40, 0.50, 0.60, 0.70\}\) confirms that the test AUC varies by only 0.035 and the test recall remains above 0.88 across the full \(\pm 40\%\) perturbation range (Figs.~S24--S26), demonstrating robustness to prior misspecification. Optimization used Adam with learning rate $1 \times 10^{-3}$, weight decay $1 \times 10^{-4}$, gradient clipping ($\lVert g \rVert_2 \leq 1.0$), and batch size 128. Training ran for a maximum of 30 epochs with early-stopping patience of 10 epochs based on training AUC and a ReduceLROnPlateau scheduler (patience 5, factor 0.5). Prior to MLP training, the 100 selected features were normalized via per-feature z-score standardization (StandardScaler fitted on training data only). Within the FTCP extraction pathways, per-channel normalization was applied to each block according to its physical quantity type (min-max for lattice parameters, coordinate normalization for atomic positions, log-scaling for reciprocal-space magnitudes).

To evaluate the framework’s robustness against label contamination—where true synthesizable phases are erroneously categorized as unlabeled—we conducted a series of stress tests by intentionally hiding up to 20\% of confirmed training positives. The pipeline’s ranking performance (AUC) remains highly stable under these conditions. Furthermore, the model exhibits a capacity for "spontaneous recovery," identifying hidden structural patterns and assigning high synthesizability probabilities to contaminated samples despite the absence of positive training labels.

\subsection*{Threshold calibration}
Standard binary classification ($p \geq 0.5$) proved inadequate under severe imbalance. We therefore implemented adaptive thresholds:
\begin{itemize}
	\item \textbf{Dual thresholds:} $p \geq 0.30 \Rightarrow$ synthesizable, $p \leq 0.25 \Rightarrow$ non-synthesizable, else uncertain.
	\item \textbf{Triple thresholds:} $p \geq 0.70 \Rightarrow$ highly synthesizable; $0.40 \leq p < 0.70 \Rightarrow$ likely synthesizable; $0.35 \leq p < 0.40 \Rightarrow$ uncertain; $p < 0.35 \Rightarrow$ non-synthesizable.
\end{itemize}
Threshold values were calibrated exclusively on the validation set via exhaustive grid search over candidate thresholds, maximizing a composite objective defined as \(0.6 \times \mathrm{recall} + 0.2 \times \mathrm{precision} + 0.2 \times \mathrm{coverage}\), subject to hard constraints requiring \(\mathrm{recall} \geq 90\%\) and \(\mathrm{coverage} \geq 80\%\) for the dual configuration and \(\mathrm{recall} \geq 80\%\) for the triple configuration. The asymmetric weighting reflects the materials-screening priority of minimizing missed synthesizable compounds. No test data were used during threshold selection. Threshold selection optimized for high recall while providing explicit uncertainty quantification, critical for applications where missing synthesizable materials incurs greater costs than investigating non-synthesizable candidates.

\subsection*{Performance evaluation}
Model performance was assessed using standard classification metrics adapted for PU learning contexts. Key metrics included:
\begin{itemize}
	\item \textbf{Area under ROC curve (AUC):} 
	\begin{equation}
		TPR=\frac{TP}{TP+FN},  FPR=\frac{FP}{FP+TN}
	\end{equation}

	Then, 
	\begin{equation}
		AUC=\int_{0}^{1}TPR(FPR)d(FPR)
	\end{equation}

	\item \textbf{Recall:} 
	\begin{equation}
		Recall=\frac{TP}{TP+FN}
	\end{equation}
	
	\item \textbf{Precision:} 
	\begin{equation}
		Precision=\frac{TP}{TP+FP}
	\end{equation}
	
	\item \textbf{Area under Precision-Recall curve (AUPRC):} 
	\begin{equation}
  		AUPRC = \sum_{n} \bigl(Recall_n - Recall_{n-1}\bigr)\cdot Precision_n
	\end{equation}

	\item \textbf{Coverage:} Coverage is relevant for dual or triple threshold strategies, where some predictions are left uncertain.
	If N is the total number of samples and $N_c$ is the number of samples that receive a confident prediction (either positive or negative), then:
	\begin{equation}
		Coverage=\frac{N_c}{N}
	\end{equation}
	
\end{itemize}

\section*{Data availability}
All crystal structures analysed in this study were obtained from the Materials Project for the years 2011--2025, and positive labels were assigned by cross-referencing Materials Project entries with the ICSD. The exact list of Materials Project identifiers used in this work are provided in \url{https://github.com/INQUIRELAB/SyntheFormer}.

\section*{Code availability}
All code for dataset construction (including MP-ICSD cross-referencing and temporal splitting), model training/inference is openly available at \url{https://github.com/INQUIRELAB/SyntheFormer}.

\section*{Acknowledgements}
This work was supported in part by the U.S. National Science Foundation under Award No. 2432082 and the Air Force Office of Scientific Research under Award No. FA9550-25-1-0322. Any opinions, findings, and conclusions or recommendations are those of the authors and do not necessarily reflect the views of the funding agencies.

\section*{Author Contributions}
D.E. conceived and implemented the SyntheFormer architecture, developed the structure-aware feature extraction and threshold calibration modules, performed data curation and model training, and designed the Fourier-Transformed Crystal Properties (FTCP) representation. SS contributed to project supervision, including materials data preprocessing and interpreting the model's chemical and physical implications. Y.M.B. conceptualized and led the project, provided overall supervision, guided the methodological framework and validation strategy, and coordinated the integration of AI and materials science components. All authors discussed the results, contributed to manuscript writing and review, and approved the final version of the paper.

\section*{Competing interests}
The authors declare that a patent application related to the methods described in this work has been filed by University of Oklahoma. All authors are listed as inventors on this application (U.S. patent application no. 19/542,123, currently pending), which covers the methodology and framework presented in this manuscript.

\section*{Supplementary information}
The supplementary file has been attached.

Correspondence and requests for materials should be addressed to Yaser M. Banad.


\begin{thebibliography}{52}
	\ifx \bisbn   \undefined \def \bisbn  #1{ISBN #1}\fi
	\ifx \binits  \undefined \def \binits#1{#1}\fi
	\ifx \bauthor  \undefined \def \bauthor#1{#1}\fi
	\ifx \batitle  \undefined \def \batitle#1{#1}\fi
	\ifx \bjtitle  \undefined \def \bjtitle#1{#1}\fi
	\ifx \bvolume  \undefined \def \bvolume#1{\textbf{#1}}\fi
	\ifx \byear  \undefined \def \byear#1{#1}\fi
	\ifx \bissue  \undefined \def \bissue#1{#1}\fi
	\ifx \bfpage  \undefined \def \bfpage#1{#1}\fi
	\ifx \blpage  \undefined \def \blpage #1{#1}\fi
	\ifx \burl  \undefined \def \burl#1{\textsf{#1}}\fi
	\ifx \doiurl  \undefined \def \doiurl#1{\url{https://doi.org/#1}}\fi
	\ifx \betal  \undefined \def \betal{\textit{et al.}}\fi
	\ifx \binstitute  \undefined \def \binstitute#1{#1}\fi
	\ifx \binstitutionaled  \undefined \def \binstitutionaled#1{#1}\fi
	\ifx \bctitle  \undefined \def \bctitle#1{#1}\fi
	\ifx \beditor  \undefined \def \beditor#1{#1}\fi
	\ifx \bpublisher  \undefined \def \bpublisher#1{#1}\fi
	\ifx \bbtitle  \undefined \def \bbtitle#1{#1}\fi
	\ifx \bedition  \undefined \def \bedition#1{#1}\fi
	\ifx \bseriesno  \undefined \def \bseriesno#1{#1}\fi
	\ifx \blocation  \undefined \def \blocation#1{#1}\fi
	\ifx \bsertitle  \undefined \def \bsertitle#1{#1}\fi
	\ifx \bsnm \undefined \def \bsnm#1{#1}\fi
	\ifx \bsuffix \undefined \def \bsuffix#1{#1}\fi
	\ifx \bparticle \undefined \def \bparticle#1{#1}\fi
	\ifx \barticle \undefined \def \barticle#1{#1}\fi
	\bibcommenthead
	\ifx \bconfdate \undefined \def \bconfdate #1{#1}\fi
	\ifx \botherref \undefined \def \botherref #1{#1}\fi
	\ifx \url \undefined \def \url#1{\textsf{#1}}\fi
	\ifx \bchapter \undefined \def \bchapter#1{#1}\fi
	\ifx \bbook \undefined \def \bbook#1{#1}\fi
	\ifx \bcomment \undefined \def \bcomment#1{#1}\fi
	\ifx \oauthor \undefined \def \oauthor#1{#1}\fi
	\ifx \citeauthoryear \undefined \def \citeauthoryear#1{#1}\fi
	\ifx \endbibitem  \undefined \def \endbibitem {}\fi
	\ifx \bconflocation  \undefined \def \bconflocation#1{#1}\fi
	\ifx \arxivurl  \undefined \def \arxivurl#1{\textsf{#1}}\fi
	\csname PreBibitemsHook\endcsname
	
	\bibitem[\protect\citeauthoryear{Bartel}{2022}]{bartel2022review}
	\begin{barticle}
		\bauthor{\bsnm{Bartel}, \binits{C.J.}}:
		\batitle{Review of computational approaches to predict the thermodynamic
			stability of inorganic solids}.
		\bjtitle{Journal of Materials Science}
		\bvolume{57}(\bissue{23}),
		\bfpage{10475}--\blpage{10498}
		(\byear{2022})
	\end{barticle}
	\endbibitem
	
	\bibitem[\protect\citeauthoryear{Pearson}{1901}]{pearson1901liii}
	\begin{barticle}
		\bauthor{\bsnm{Pearson}, \binits{K.}}:
		\batitle{Liii. on lines and planes of closest fit to systems of points in
			space}.
		\bjtitle{The London, Edinburgh, and Dublin philosophical magazine and journal
			of science}
		\bvolume{2}(\bissue{11}),
		\bfpage{559}--\blpage{572}
		(\byear{1901})
	\end{barticle}
	\endbibitem
	
	\bibitem[\protect\citeauthoryear{Kajita et~al.}{2017}]{kajita2017universal}
	\begin{barticle}
		\bauthor{\bsnm{Kajita}, \binits{S.}},
		\bauthor{\bsnm{Ohba}, \binits{N.}},
		\bauthor{\bsnm{Jinnouchi}, \binits{R.}},
		\bauthor{\bsnm{Asahi}, \binits{R.}}:
		\batitle{A universal 3d voxel descriptor for solid-state material informatics
			with deep convolutional neural networks}.
		\bjtitle{Scientific reports}
		\bvolume{7}(\bissue{1}),
		\bfpage{16991}
		(\byear{2017})
	\end{barticle}
	\endbibitem
	
	\bibitem[\protect\citeauthoryear{Korolev et~al.}{2020}]{korolev2020machine}
	\begin{barticle}
		\bauthor{\bsnm{Korolev}, \binits{V.}},
		\bauthor{\bsnm{Mitrofanov}, \binits{A.}},
		\bauthor{\bsnm{Eliseev}, \binits{A.}},
		\bauthor{\bsnm{Tkachenko}, \binits{V.}}:
		\batitle{Machine-learning-assisted search for functional materials over
			extended chemical space}.
		\bjtitle{Materials Horizons}
		\bvolume{7}(\bissue{10}),
		\bfpage{2710}--\blpage{2718}
		(\byear{2020})
	\end{barticle}
	\endbibitem
	
	\bibitem[\protect\citeauthoryear{Zagorac et~al.}{2019}]{zagorac2019recent}
	\begin{barticle}
		\bauthor{\bsnm{Zagorac}, \binits{D.}},
		\bauthor{\bsnm{M{\"u}ller}, \binits{H.}},
		\bauthor{\bsnm{Ruehl}, \binits{S.}},
		\bauthor{\bsnm{Zagorac}, \binits{J.}},
		\bauthor{\bsnm{Rehme}, \binits{S.}}:
		\batitle{Recent developments in the inorganic crystal structure database:
			theoretical crystal structure data and related features}.
		\bjtitle{Applied Crystallography}
		\bvolume{52}(\bissue{5}),
		\bfpage{918}--\blpage{925}
		(\byear{2019})
	\end{barticle}
	\endbibitem
	
	\bibitem[\protect\citeauthoryear{Xie and Grossman}{2018}]{xie2018crystal}
	\begin{barticle}
		\bauthor{\bsnm{Xie}, \binits{T.}},
		\bauthor{\bsnm{Grossman}, \binits{J.C.}}:
		\batitle{Crystal graph convolutional neural networks for an accurate and
			interpretable prediction of material properties}.
		\bjtitle{Physical review letters}
		\bvolume{120}(\bissue{14}),
		\bfpage{145301}
		(\byear{2018})
	\end{barticle}
	\endbibitem
	
	\bibitem[\protect\citeauthoryear{Qiu et~al.}{2025}]{qiu2025massive}
	\begin{barticle}
		\bauthor{\bsnm{Qiu}, \binits{Z.}},
		\bauthor{\bsnm{Jin}, \binits{L.}},
		\bauthor{\bsnm{Du}, \binits{Z.}},
		\bauthor{\bsnm{Chen}, \binits{H.}},
		\bauthor{\bsnm{Mao}, \binits{G.}},
		\bauthor{\bsnm{Cen}, \binits{Y.}},
		\bauthor{\bsnm{Sun}, \binits{S.}},
		\bauthor{\bsnm{Mei}, \binits{Y.}},
		\bauthor{\bsnm{Zhang}, \binits{H.}}:
		\batitle{Massive discovery of crystal structures across dimensionalities by
			leveraging vector quantization}.
		\bjtitle{npj Computational Materials}
		\bvolume{11}(\bissue{1}),
		\bfpage{184}
		(\byear{2025})
	\end{barticle}
	\endbibitem
	
	\bibitem[\protect\citeauthoryear{Park et~al.}{2024}]{park2024has}
	\begin{barticle}
		\bauthor{\bsnm{Park}, \binits{H.}},
		\bauthor{\bsnm{Li}, \binits{Z.}},
		\bauthor{\bsnm{Walsh}, \binits{A.}}:
		\batitle{Has generative artificial intelligence solved inverse materials
			design?}
		\bjtitle{Matter}
		\bvolume{7}(\bissue{7}),
		\bfpage{2355}--\blpage{2367}
		(\byear{2024})
	\end{barticle}
	\endbibitem
	
	\bibitem[\protect\citeauthoryear{Kohn and Sham}{1965}]{kohn1965self}
	\begin{barticle}
		\bauthor{\bsnm{Kohn}, \binits{W.}},
		\bauthor{\bsnm{Sham}, \binits{L.J.}}:
		\batitle{Self-consistent equations including exchange and correlation effects}.
		\bjtitle{Physical review}
		\bvolume{140}(\bissue{4A}),
		\bfpage{1133}
		(\byear{1965})
	\end{barticle}
	\endbibitem
	
	\bibitem[\protect\citeauthoryear{Jain et~al.}{2013}]{jain2013commentary}
	\begin{botherref}
		\oauthor{\bsnm{Jain}, \binits{A.}},
		\oauthor{\bsnm{Ong}, \binits{S.P.}},
		\oauthor{\bsnm{Hautier}, \binits{G.}},
		\oauthor{\bsnm{Chen}, \binits{W.}},
		\oauthor{\bsnm{Richards}, \binits{W.D.}},
		\oauthor{\bsnm{Dacek}, \binits{S.}},
		\oauthor{\bsnm{Cholia}, \binits{S.}},
		\oauthor{\bsnm{Gunter}, \binits{D.}},
		\oauthor{\bsnm{Skinner}, \binits{D.}},
		\oauthor{\bsnm{Ceder}, \binits{G.}}, et al.:
		Commentary: The materials project: A materials genome approach to accelerating
		materials innovation.
		APL materials
		\textbf{1}(1)
		(2013)
	\end{botherref}
	\endbibitem
	
	\bibitem[\protect\citeauthoryear{Kirklin et~al.}{2015}]{kirklin2015open}
	\begin{barticle}
		\bauthor{\bsnm{Kirklin}, \binits{S.}},
		\bauthor{\bsnm{Saal}, \binits{J.E.}},
		\bauthor{\bsnm{Meredig}, \binits{B.}},
		\bauthor{\bsnm{Thompson}, \binits{A.}},
		\bauthor{\bsnm{Doak}, \binits{J.W.}},
		\bauthor{\bsnm{Aykol}, \binits{M.}},
		\bauthor{\bsnm{R{\"u}hl}, \binits{S.}},
		\bauthor{\bsnm{Wolverton}, \binits{C.}}:
		\batitle{The open quantum materials database (oqmd): assessing the accuracy of
			dft formation energies}.
		\bjtitle{npj Computational Materials}
		\bvolume{1}(\bissue{1}),
		\bfpage{1}--\blpage{15}
		(\byear{2015})
	\end{barticle}
	\endbibitem
	
	\bibitem[\protect\citeauthoryear{Curtarolo et~al.}{2012}]{curtarolo2012aflow}
	\begin{barticle}
		\bauthor{\bsnm{Curtarolo}, \binits{S.}},
		\bauthor{\bsnm{Setyawan}, \binits{W.}},
		\bauthor{\bsnm{Hart}, \binits{G.L.}},
		\bauthor{\bsnm{Jahnatek}, \binits{M.}},
		\bauthor{\bsnm{Chepulskii}, \binits{R.V.}},
		\bauthor{\bsnm{Taylor}, \binits{R.H.}},
		\bauthor{\bsnm{Wang}, \binits{S.}},
		\bauthor{\bsnm{Xue}, \binits{J.}},
		\bauthor{\bsnm{Yang}, \binits{K.}},
		\bauthor{\bsnm{Levy}, \binits{O.}}, \betal:
		\batitle{Aflow: An automatic framework for high-throughput materials
			discovery}.
		\bjtitle{Computational Materials Science}
		\bvolume{58},
		\bfpage{218}--\blpage{226}
		(\byear{2012})
	\end{barticle}
	\endbibitem
	
	\bibitem[\protect\citeauthoryear{Sun et~al.}{2016}]{sun2016thermodynamic}
	\begin{barticle}
		\bauthor{\bsnm{Sun}, \binits{W.}},
		\bauthor{\bsnm{Dacek}, \binits{S.T.}},
		\bauthor{\bsnm{Ong}, \binits{S.P.}},
		\bauthor{\bsnm{Hautier}, \binits{G.}},
		\bauthor{\bsnm{Jain}, \binits{A.}},
		\bauthor{\bsnm{Richards}, \binits{W.D.}},
		\bauthor{\bsnm{Gamst}, \binits{A.C.}},
		\bauthor{\bsnm{Persson}, \binits{K.A.}},
		\bauthor{\bsnm{Ceder}, \binits{G.}}:
		\batitle{The thermodynamic scale of inorganic crystalline metastability}.
		\bjtitle{Science advances}
		\bvolume{2}(\bissue{11}),
		\bfpage{1600225}
		(\byear{2016})
	\end{barticle}
	\endbibitem
	
	\bibitem[\protect\citeauthoryear{Lee et~al.}{2022}]{lee2022machine}
	\begin{barticle}
		\bauthor{\bsnm{Lee}, \binits{A.}},
		\bauthor{\bsnm{Sarker}, \binits{S.}},
		\bauthor{\bsnm{Saal}, \binits{J.E.}},
		\bauthor{\bsnm{Ward}, \binits{L.}},
		\bauthor{\bsnm{Borg}, \binits{C.}},
		\bauthor{\bsnm{Mehta}, \binits{A.}},
		\bauthor{\bsnm{Wolverton}, \binits{C.}}:
		\batitle{Machine learned synthesizability predictions aided by density
			functional theory}.
		\bjtitle{Communications Materials}
		\bvolume{3}(\bissue{1}),
		\bfpage{73}
		(\byear{2022})
	\end{barticle}
	\endbibitem
	
	\bibitem[\protect\citeauthoryear{Yu et~al.}{2024}]{yu2024far}
	\begin{barticle}
		\bauthor{\bsnm{Yu}, \binits{Y.}},
		\bauthor{\bsnm{Qin}, \binits{Z.}},
		\bauthor{\bsnm{Zhang}, \binits{X.}},
		\bauthor{\bsnm{Chen}, \binits{Y.}},
		\bauthor{\bsnm{Qin}, \binits{G.}},
		\bauthor{\bsnm{Li}, \binits{S.}}:
		\batitle{Far-from-equilibrium processing opens kinetic paths for engineering
			novel materials by breaking thermodynamic limits}.
		\bjtitle{ACS Materials Letters}
		\bvolume{7}(\bissue{1}),
		\bfpage{319}--\blpage{332}
		(\byear{2024})
	\end{barticle}
	\endbibitem
	
	\bibitem[\protect\citeauthoryear{Kim et~al.}{2025}]{kim2025explainable}
	\begin{barticle}
		\bauthor{\bsnm{Kim}, \binits{S.}},
		\bauthor{\bsnm{Schrier}, \binits{J.}},
		\bauthor{\bsnm{Jung}, \binits{Y.}}:
		\batitle{Explainable synthesizability prediction of inorganic crystal
			polymorphs using large language models}.
		\bjtitle{Angewandte Chemie International Edition}
		\bvolume{64}(\bissue{19}),
		\bfpage{202423950}
		(\byear{2025})
	\end{barticle}
	\endbibitem
	
	\bibitem[\protect\citeauthoryear{Song et~al.}{2025}]{song2025accurate}
	\begin{barticle}
		\bauthor{\bsnm{Song}, \binits{Z.}},
		\bauthor{\bsnm{Lu}, \binits{S.}},
		\bauthor{\bsnm{Ju}, \binits{M.}},
		\bauthor{\bsnm{Zhou}, \binits{Q.}},
		\bauthor{\bsnm{Wang}, \binits{J.}}:
		\batitle{Accurate prediction of synthesizability and precursors of 3d crystal
			structures via large language models}.
		\bjtitle{Nature Communications}
		\bvolume{16}(\bissue{1}),
		\bfpage{6530}
		(\byear{2025})
	\end{barticle}
	\endbibitem
	
	\bibitem[\protect\citeauthoryear{Sokolikova and
		Mattevi}{2020}]{sokolikova2020direct}
	\begin{barticle}
		\bauthor{\bsnm{Sokolikova}, \binits{M.S.}},
		\bauthor{\bsnm{Mattevi}, \binits{C.}}:
		\batitle{Direct synthesis of metastable phases of 2d transition metal
			dichalcogenides}.
		\bjtitle{Chemical Society Reviews}
		\bvolume{49}(\bissue{12}),
		\bfpage{3952}--\blpage{3980}
		(\byear{2020})
	\end{barticle}
	\endbibitem
	
	\bibitem[\protect\citeauthoryear{Ye et~al.}{2018}]{ye2018deep}
	\begin{barticle}
		\bauthor{\bsnm{Ye}, \binits{W.}},
		\bauthor{\bsnm{Chen}, \binits{C.}},
		\bauthor{\bsnm{Wang}, \binits{Z.}},
		\bauthor{\bsnm{Chu}, \binits{I.-H.}},
		\bauthor{\bsnm{Ong}, \binits{S.P.}}:
		\batitle{Deep neural networks for accurate predictions of crystal stability}.
		\bjtitle{Nature communications}
		\bvolume{9}(\bissue{1}),
		\bfpage{3800}
		(\byear{2018})
	\end{barticle}
	\endbibitem
	
	\bibitem[\protect\citeauthoryear{Davariashtiyani
		et~al.}{2021}]{davariashtiyani2021predicting}
	\begin{barticle}
		\bauthor{\bsnm{Davariashtiyani}, \binits{A.}},
		\bauthor{\bsnm{Kadkhodaie}, \binits{Z.}},
		\bauthor{\bsnm{Kadkhodaei}, \binits{S.}}:
		\batitle{Predicting synthesizability of crystalline materials via deep
			learning}.
		\bjtitle{Communications Materials}
		\bvolume{2}(\bissue{1}),
		\bfpage{115}
		(\byear{2021})
	\end{barticle}
	\endbibitem
	
	\bibitem[\protect\citeauthoryear{Kononova et~al.}{2019}]{kononova2019text}
	\begin{barticle}
		\bauthor{\bsnm{Kononova}, \binits{O.}},
		\bauthor{\bsnm{Huo}, \binits{H.}},
		\bauthor{\bsnm{He}, \binits{T.}},
		\bauthor{\bsnm{Rong}, \binits{Z.}},
		\bauthor{\bsnm{Botari}, \binits{T.}},
		\bauthor{\bsnm{Sun}, \binits{W.}},
		\bauthor{\bsnm{Tshitoyan}, \binits{V.}},
		\bauthor{\bsnm{Ceder}, \binits{G.}}:
		\batitle{Text-mined dataset of inorganic materials synthesis recipes}.
		\bjtitle{Scientific data}
		\bvolume{6}(\bissue{1}),
		\bfpage{203}
		(\byear{2019})
	\end{barticle}
	\endbibitem
	
	\bibitem[\protect\citeauthoryear{Cruse et~al.}{2023}]{cruse2023text}
	\begin{barticle}
		\bauthor{\bsnm{Cruse}, \binits{K.}},
		\bauthor{\bsnm{Baibakova}, \binits{V.}},
		\bauthor{\bsnm{Abdelsamie}, \binits{M.}},
		\bauthor{\bsnm{Hong}, \binits{K.}},
		\bauthor{\bsnm{Bartel}, \binits{C.J.}},
		\bauthor{\bsnm{Trewartha}, \binits{A.}},
		\bauthor{\bsnm{Jain}, \binits{A.}},
		\bauthor{\bsnm{Sutter-Fella}, \binits{C.M.}},
		\bauthor{\bsnm{Ceder}, \binits{G.}}:
		\batitle{Text mining the literature to inform experiments and rationalize
			impurity phase formation for bifeo3}.
		\bjtitle{Chemistry of Materials}
		\bvolume{36}(\bissue{2}),
		\bfpage{772}--\blpage{785}
		(\byear{2023})
	\end{barticle}
	\endbibitem
	
	\bibitem[\protect\citeauthoryear{Bekker and Davis}{2020}]{bekker2020learning}
	\begin{barticle}
		\bauthor{\bsnm{Bekker}, \binits{J.}},
		\bauthor{\bsnm{Davis}, \binits{J.}}:
		\batitle{Learning from positive and unlabeled data: A survey}.
		\bjtitle{Machine learning}
		\bvolume{109}(\bissue{4}),
		\bfpage{719}--\blpage{760}
		(\byear{2020})
	\end{barticle}
	\endbibitem
	
	\bibitem[\protect\citeauthoryear{Kiryo et~al.}{2017}]{kiryo2017positive}
	\begin{botherref}
		\oauthor{\bsnm{Kiryo}, \binits{R.}},
		\oauthor{\bsnm{Niu}, \binits{G.}},
		\oauthor{\bsnm{Du~Plessis}, \binits{M.C.}},
		\oauthor{\bsnm{Sugiyama}, \binits{M.}}:
		Positive-unlabeled learning with non-negative risk estimator.
		Advances in neural information processing systems
		\textbf{30}
		(2017)
	\end{botherref}
	\endbibitem
	
	\bibitem[\protect\citeauthoryear{Elkan and Noto}{2008}]{elkan2008learning}
	\begin{bchapter}
		\bauthor{\bsnm{Elkan}, \binits{C.}},
		\bauthor{\bsnm{Noto}, \binits{K.}}:
		\bctitle{Learning classifiers from only positive and unlabeled data}.
		In: \bbtitle{Proceedings of the 14th ACM SIGKDD International Conference on
			Knowledge Discovery and Data Mining},
		pp. \bfpage{213}--\blpage{220}
		(\byear{2008})
	\end{bchapter}
	\endbibitem
	
	\bibitem[\protect\citeauthoryear{Chung et~al.}{2025}]{chung2025solid}
	\begin{botherref}
		\oauthor{\bsnm{Chung}, \binits{V.}},
		\oauthor{\bsnm{Walsh}, \binits{A.}},
		\oauthor{\bsnm{Payne}, \binits{D.J.}}:
		Solid-state synthesizability predictions using positive-unlabeled learning from
		human-curated literature data.
		Digital Discovery
		(2025)
	\end{botherref}
	\endbibitem
	
	\bibitem[\protect\citeauthoryear{Levin}{2018}]{levin2018NIST}
	\begin{barticle}
		\bauthor{\bsnm{Levin}, \binits{I.}}:
		\batitle{Nist inorganic crystal structure database (icsd)}.
		\bjtitle{National Institute of Standards and Technology}
		(\byear{2018})
		\doiurl{10.18434/M32147}
	\end{barticle}
	\endbibitem
	
	\bibitem[\protect\citeauthoryear{Antoniuk
		et~al.}{2023}]{antoniuk2023predicting}
	\begin{barticle}
		\bauthor{\bsnm{Antoniuk}, \binits{E.R.}},
		\bauthor{\bsnm{Cheon}, \binits{G.}},
		\bauthor{\bsnm{Wang}, \binits{G.}},
		\bauthor{\bsnm{Bernstein}, \binits{D.}},
		\bauthor{\bsnm{Cai}, \binits{W.}},
		\bauthor{\bsnm{Reed}, \binits{E.J.}}:
		\batitle{Predicting the synthesizability of crystalline inorganic materials
			from the data of known material compositions}.
		\bjtitle{npj Computational Materials}
		\bvolume{9}(\bissue{1}),
		\bfpage{155}
		(\byear{2023})
	\end{barticle}
	\endbibitem
	
	\bibitem[\protect\citeauthoryear{Zhu et~al.}{2023}]{zhu2023predicting}
	\begin{barticle}
		\bauthor{\bsnm{Zhu}, \binits{R.}},
		\bauthor{\bsnm{Tian}, \binits{S.I.P.}},
		\bauthor{\bsnm{Ren}, \binits{Z.}},
		\bauthor{\bsnm{Li}, \binits{J.}},
		\bauthor{\bsnm{Buonassisi}, \binits{T.}},
		\bauthor{\bsnm{Hippalgaonkar}, \binits{K.}}:
		\batitle{Predicting synthesizability using machine learning on databases of
			existing inorganic materials}.
		\bjtitle{ACS omega}
		\bvolume{8}(\bissue{9}),
		\bfpage{8210}--\blpage{8218}
		(\byear{2023})
	\end{barticle}
	\endbibitem
	
	\bibitem[\protect\citeauthoryear{Jang et~al.}{2024}]{jang2024synthesizability}
	\begin{barticle}
		\bauthor{\bsnm{Jang}, \binits{J.}},
		\bauthor{\bsnm{Noh}, \binits{J.}},
		\bauthor{\bsnm{Zhou}, \binits{L.}},
		\bauthor{\bsnm{Gu}, \binits{G.H.}},
		\bauthor{\bsnm{Gregoire}, \binits{J.M.}},
		\bauthor{\bsnm{Jung}, \binits{Y.}}:
		\batitle{Synthesizability of materials stoichiometry using semi-supervised
			learning}.
		\bjtitle{Matter}
		\bvolume{7}(\bissue{6}),
		\bfpage{2294}--\blpage{2312}
		(\byear{2024})
	\end{barticle}
	\endbibitem
	
	\bibitem[\protect\citeauthoryear{Jang et~al.}{2020}]{jang2020structure}
	\begin{barticle}
		\bauthor{\bsnm{Jang}, \binits{J.}},
		\bauthor{\bsnm{Gu}, \binits{G.H.}},
		\bauthor{\bsnm{Noh}, \binits{J.}},
		\bauthor{\bsnm{Kim}, \binits{J.}},
		\bauthor{\bsnm{Jung}, \binits{Y.}}:
		\batitle{Structure-based synthesizability prediction of crystals using
			partially supervised learning}.
		\bjtitle{Journal of the American Chemical Society}
		\bvolume{142}(\bissue{44}),
		\bfpage{18836}--\blpage{18843}
		(\byear{2020})
	\end{barticle}
	\endbibitem
	
	\bibitem[\protect\citeauthoryear{Gu et~al.}{2022}]{gu2022perovskite}
	\begin{barticle}
		\bauthor{\bsnm{Gu}, \binits{G.H.}},
		\bauthor{\bsnm{Jang}, \binits{J.}},
		\bauthor{\bsnm{Noh}, \binits{J.}},
		\bauthor{\bsnm{Walsh}, \binits{A.}},
		\bauthor{\bsnm{Jung}, \binits{Y.}}:
		\batitle{Perovskite synthesizability using graph neural networks}.
		\bjtitle{npj Computational Materials}
		\bvolume{8}(\bissue{1}),
		\bfpage{71}
		(\byear{2022})
	\end{barticle}
	\endbibitem
	
	\bibitem[\protect\citeauthoryear{Amariamir et~al.}{2025}]{amariamir2025syncotrain}
	\begin{barticle}
		\bauthor{\bsnm{Amariamir}, \binits{S.}},
		\bauthor{\bsnm{George}, \binits{J.}},
		\bauthor{\bsnm{Benner}, \binits{P.}}:
		\batitle{SynCoTrain: a dual classifier PU-learning framework for synthesizability prediction}.
		\bjtitle{Digital Discovery}
		\bvolume{4}(\bissue{6}),
		\bfpage{1437}--\blpage{1448}
		(\byear{2025})
	\end{barticle}
	\endbibitem

	\bibitem[\protect\citeauthoryear{Mordelet and Vert}{2014}]{mordelet2014bagging}
	\begin{barticle}
		\bauthor{\bsnm{Mordelet}, \binits{F.}},
		\bauthor{\bsnm{Vert}, \binits{J.-P.}}:
		\batitle{A bagging svm to learn from positive and unlabeled examples}.
		\bjtitle{Pattern Recognition Letters}
		\bvolume{37},
		\bfpage{201}--\blpage{209}
		(\byear{2014})
	\end{barticle}
	\endbibitem

	\bibitem[\protect\citeauthoryear{Chen et~al.}{2019}]{chen2019graph}
	\begin{barticle}
		\bauthor{\bsnm{Chen}, \binits{C.}},
		\bauthor{\bsnm{Ye}, \binits{W.}},
		\bauthor{\bsnm{Zuo}, \binits{Y.}},
		\bauthor{\bsnm{Zheng}, \binits{C.}},
		\bauthor{\bsnm{Ong}, \binits{S.P.}}:
		\batitle{Graph networks as a universal machine learning framework for molecules and crystals}.
		\bjtitle{Chemistry of Materials}
		\bvolume{31}(\bissue{9}),
		\bfpage{3564}--\blpage{3572}
		(\byear{2019})
	\end{barticle}
	\endbibitem
	
	\bibitem[\protect\citeauthoryear{Ren et~al.}{2022}]{ren2022invertible}
	\begin{barticle}
		\bauthor{\bsnm{Ren}, \binits{Z.}},
		\bauthor{\bsnm{Tian}, \binits{S.I.P.}},
		\bauthor{\bsnm{Noh}, \binits{J.}},
		\bauthor{\bsnm{Oviedo}, \binits{F.}},
		\bauthor{\bsnm{Xing}, \binits{G.}},
		\bauthor{\bsnm{Li}, \binits{J.}},
		\bauthor{\bsnm{Liang}, \binits{Q.}},
		\bauthor{\bsnm{Zhu}, \binits{R.}},
		\bauthor{\bsnm{Aberle}, \binits{A.G.}},
		\bauthor{\bsnm{Sun}, \binits{S.}}, \betal:
		\batitle{An invertible crystallographic representation for general inverse
			design of inorganic crystals with targeted properties}.
		\bjtitle{Matter}
		\bvolume{5}(\bissue{1}),
		\bfpage{314}--\blpage{335}
		(\byear{2022})
	\end{barticle}
	\endbibitem
	
	\bibitem[\protect\citeauthoryear{Du et~al.}{2024}]{du2024ctgnn}
	\begin{botherref}
		\oauthor{\bsnm{Du}, \binits{Z.}},
		\oauthor{\bsnm{Jin}, \binits{L.}},
		\oauthor{\bsnm{Shu}, \binits{L.}},
		\oauthor{\bsnm{Cen}, \binits{Y.}},
		\oauthor{\bsnm{Xu}, \binits{Y.}},
		\oauthor{\bsnm{Mei}, \binits{Y.}},
		\oauthor{\bsnm{Zhang}, \binits{H.}}:
		Ctgnn: Crystal transformer graph neural network for crystal material property
		prediction.
		arXiv preprint arXiv:2405.11502
		(2024)
	\end{botherref}
	\endbibitem
	
	\bibitem[\protect\citeauthoryear{Vaswani et~al.}{2017}]{vaswani2017attention}
	\begin{botherref}
		\oauthor{\bsnm{Vaswani}, \binits{A.}},
		\oauthor{\bsnm{Shazeer}, \binits{N.}},
		\oauthor{\bsnm{Parmar}, \binits{N.}},
		\oauthor{\bsnm{Uszkoreit}, \binits{J.}},
		\oauthor{\bsnm{Jones}, \binits{L.}},
		\oauthor{\bsnm{Gomez}, \binits{A.N.}},
		\oauthor{\bsnm{Kaiser}, \binits{{\L}.}},
		\oauthor{\bsnm{Polosukhin}, \binits{I.}}:
		Attention is all you need.
		Advances in neural information processing systems
		\textbf{30}
		(2017)
	\end{botherref}
	\endbibitem
	
	\bibitem[\protect\citeauthoryear{Ziletti et~al.}{2018}]{ziletti2018insightful}
	\begin{barticle}
		\bauthor{\bsnm{Ziletti}, \binits{A.}},
		\bauthor{\bsnm{Kumar}, \binits{D.}},
		\bauthor{\bsnm{Scheffler}, \binits{M.}},
		\bauthor{\bsnm{Ghiringhelli}, \binits{L.M.}}:
		\batitle{Insightful classification of crystal structures using deep learning}.
		\bjtitle{Nature communications}
		\bvolume{9}(\bissue{1}),
		\bfpage{2775}
		(\byear{2018})
	\end{barticle}
	\endbibitem
	
	\bibitem[\protect\citeauthoryear{Taniai et~al.}{2024}]{taniai2024crystalformer}
	\begin{botherref}
		\oauthor{\bsnm{Taniai}, \binits{T.}},
		\oauthor{\bsnm{Igarashi}, \binits{R.}},
		\oauthor{\bsnm{Suzuki}, \binits{Y.}},
		\oauthor{\bsnm{Chiba}, \binits{N.}},
		\oauthor{\bsnm{Saito}, \binits{K.}},
		\oauthor{\bsnm{Ushiku}, \binits{Y.}},
		\oauthor{\bsnm{Ono}, \binits{K.}}:
		Crystalformer: Infinitely connected attention for periodic structure encoding.
		arXiv preprint arXiv:2403.11686
		(2024)
	\end{botherref}
	\endbibitem
	
	\bibitem[\protect\citeauthoryear{Kim et~al.}{2024}]{kim2024predicting}
	\begin{barticle}
		\bauthor{\bsnm{Kim}, \binits{S.}},
		\bauthor{\bsnm{Noh}, \binits{J.}},
		\bauthor{\bsnm{Gu}, \binits{G.H.}},
		\bauthor{\bsnm{Chen}, \binits{S.}},
		\bauthor{\bsnm{Jung}, \binits{Y.}}:
		\batitle{Predicting synthesis recipes of inorganic crystal materials using
			elementwise template formulation}.
		\bjtitle{Chemical Science}
		\bvolume{15}(\bissue{3}),
		\bfpage{1039}--\blpage{1045}
		(\byear{2024})
	\end{barticle}
	\endbibitem
	
	\bibitem[\protect\citeauthoryear{Sahoo et~al.}{2021}]{sahoo2021reliable}
	\begin{barticle}
		\bauthor{\bsnm{Sahoo}, \binits{R.}},
		\bauthor{\bsnm{Zhao}, \binits{S.}},
		\bauthor{\bsnm{Chen}, \binits{A.}},
		\bauthor{\bsnm{Ermon}, \binits{S.}}:
		\batitle{Reliable decisions with threshold calibration}.
		\bjtitle{Advances in Neural Information Processing Systems}
		\bvolume{34},
		\bfpage{1831}--\blpage{1844}
		(\byear{2021})
	\end{barticle}
	\endbibitem
	
	\bibitem[\protect\citeauthoryear{Ebrahimzadeh
		et~al.}{2025}]{ebrahimzadeh2025accelerated}
	\begin{barticle}
		\bauthor{\bsnm{Ebrahimzadeh}, \binits{D.}},
		\bauthor{\bsnm{Sharif}, \binits{S.S.}},
		\bauthor{\bsnm{Banad}, \binits{Y.M.}}:
		\batitle{Accelerated discovery of vanadium oxide compositions: A wgan-vae
			framework for materials design}.
		\bjtitle{Materials Today Electronics}
		\bvolume{13},
		\bfpage{100155}
		(\byear{2025})
	\end{barticle}
	\endbibitem
	
	\bibitem[\protect\citeauthoryear{Schmidt et~al.}{2021}]{schmidt2021crystal}
	\begin{barticle}
		\bauthor{\bsnm{Schmidt}, \binits{J.}},
		\bauthor{\bsnm{Pettersson}, \binits{L.}},
		\bauthor{\bsnm{Verdozzi}, \binits{C.}},
		\bauthor{\bsnm{Botti}, \binits{S.}},
		\bauthor{\bsnm{Marques}, \binits{M.A.}}:
		\batitle{Crystal graph attention networks for the prediction of stable
			materials}.
		\bjtitle{Science advances}
		\bvolume{7}(\bissue{49}),
		\bfpage{7948}
		(\byear{2021})
	\end{barticle}
	\endbibitem
	
	\bibitem[\protect\citeauthoryear{Lee et~al.}{2025}]{lee2025cast}
	\begin{botherref}
		\oauthor{\bsnm{Lee}, \binits{J.}},
		\oauthor{\bsnm{Park}, \binits{C.}},
		\oauthor{\bsnm{Yang}, \binits{H.}},
		\oauthor{\bsnm{Lim}, \binits{S.}},
		\oauthor{\bsnm{Lim}, \binits{W.}},
		\oauthor{\bsnm{Han}, \binits{S.}}:
		Cast: Cross attention based multimodal fusion of structure and text for
		materials property prediction.
		arXiv preprint arXiv:2502.06836
		(2025)
	\end{botherref}
	\endbibitem
	
	\bibitem[\protect\citeauthoryear{Breiman}{2001}]{breiman2001random}
	\begin{barticle}
		\bauthor{\bsnm{Breiman}, \binits{L.}}:
		\batitle{Random forests}.
		\bjtitle{Machine learning}
		\bvolume{45}(\bissue{1}),
		\bfpage{5}--\blpage{32}
		(\byear{2001})
	\end{barticle}
	\endbibitem
	
	\bibitem[\protect\citeauthoryear{Jolliffe and
		Cadima}{2016}]{jolliffe2016principal}
	\begin{barticle}
		\bauthor{\bsnm{Jolliffe}, \binits{I.T.}},
		\bauthor{\bsnm{Cadima}, \binits{J.}}:
		\batitle{Principal component analysis: a review and recent developments}.
		\bjtitle{Philosophical transactions of the royal society A: Mathematical,
			Physical and Engineering Sciences}
		\bvolume{374}(\bissue{2065}),
		\bfpage{20150202}
		(\byear{2016})
	\end{barticle}
	\endbibitem
	
	\bibitem[\protect\citeauthoryear{Guyon and
		Elisseeff}{2003}]{guyon2003introduction}
	\begin{barticle}
		\bauthor{\bsnm{Guyon}, \binits{I.}},
		\bauthor{\bsnm{Elisseeff}, \binits{A.}}:
		\batitle{An introduction to variable and feature selection}.
		\bjtitle{Journal of machine learning research}
		\bvolume{3}(\bissue{Mar}),
		\bfpage{1157}--\blpage{1182}
		(\byear{2003})
	\end{barticle}
	\endbibitem
	
	\bibitem[\protect\citeauthoryear{Tibshirani}{1996}]{tibshirani1996regression}
	\begin{barticle}
		\bauthor{\bsnm{Tibshirani}, \binits{R.}}:
		\batitle{Regression shrinkage and selection via the lasso}.
		\bjtitle{Journal of the Royal Statistical Society Series B: Statistical
			Methodology}
		\bvolume{58}(\bissue{1}),
		\bfpage{267}--\blpage{288}
		(\byear{1996})
	\end{barticle}
	\endbibitem
	
	\bibitem[\protect\citeauthoryear{Zou and Hastie}{2005}]{zou2005regularization}
	\begin{barticle}
		\bauthor{\bsnm{Zou}, \binits{H.}},
		\bauthor{\bsnm{Hastie}, \binits{T.}}:
		\batitle{Regularization and variable selection via the elastic net}.
		\bjtitle{Journal of the Royal Statistical Society Series B: Statistical
			Methodology}
		\bvolume{67}(\bissue{2}),
		\bfpage{301}--\blpage{320}
		(\byear{2005})
	\end{barticle}
	\endbibitem
	
	\bibitem[\protect\citeauthoryear{Kuhn and Johnson}{2019}]{kuhn2019feature}
	\begin{bbook}
		\bauthor{\bsnm{Kuhn}, \binits{M.}},
		\bauthor{\bsnm{Johnson}, \binits{K.}}:
		\bbtitle{Feature Engineering and Selection: A Practical Approach for Predictive
			Models}.
		\bpublisher{Chapman and Hall/CRC}
		(\byear{2019})
	\end{bbook}
	\endbibitem
	
	\bibitem[\protect\citeauthoryear{Kraskov et~al.}{2004}]{kraskov2004estimating}
	\begin{barticle}
		\bauthor{\bsnm{Kraskov}, \binits{A.}},
		\bauthor{\bsnm{St{\"o}gbauer}, \binits{H.}},
		\bauthor{\bsnm{Grassberger}, \binits{P.}}:
		\batitle{Estimating mutual information}.
		\bjtitle{Physical Review E--Statistical, Nonlinear, and Soft Matter Physics}
		\bvolume{69}(\bissue{6}),
		\bfpage{066138}
		(\byear{2004})
	\end{barticle}
	\endbibitem
	
	\bibitem[\protect\citeauthoryear{Peng et~al.}{2005}]{peng2005feature}
	\begin{barticle}
		\bauthor{\bsnm{Peng}, \binits{H.}},
		\bauthor{\bsnm{Long}, \binits{F.}},
		\bauthor{\bsnm{Ding}, \binits{C.}}:
		\batitle{Feature selection based on mutual information criteria of
			max-dependency, max-relevance, and min-redundancy}.
		\bjtitle{IEEE Transactions on pattern analysis and machine intelligence}
		\bvolume{27}(\bissue{8}),
		\bfpage{1226}--\blpage{1238}
		(\byear{2005})
	\end{barticle}
	\endbibitem
	
	\bibitem[\protect\citeauthoryear{Hastie}{2009}]{hastie2009elements}
	\begin{botherref}
		\oauthor{\bsnm{Hastie}, \binits{T.}}:
		The elements of statistical learning: data mining, inference, and prediction.
		Springer
		(2009)
	\end{botherref}
	\endbibitem

	\bibitem[\protect\citeauthoryear{Gamage et~al.}{2025}]{gamage2025interplay}
	\begin{barticle}
		\bauthor{\bsnm{Gamage}, \binits{E.H.}},
		\bauthor{\bsnm{Hewage}, \binits{N.W.}},
		\bauthor{\bsnm{Soto}, \binits{E.}},
		\bauthor{\bsnm{Aslam}, \binits{H.Z.}},
		\bauthor{\bsnm{Bai}, \binits{J.}},
		\bauthor{\bsnm{Alsaad}, \binits{A.}},
		\bauthor{\bsnm{Viswanathan}, \binits{G.}},
		\bauthor{\bsnm{Yox}, \binits{P.}},
		\bauthor{\bsnm{Garlea}, \binits{V.O.}},
		\bauthor{\bsnm{Kamali}, \binits{S.}}
		et~al.:
		\batitle{Interplay of 3d and 4f magnetism in chiral Y$_6$FeSi$_2$S$_{14}$ and Tb$_6$FeSi$_2$S$_{14}$ chalcogenides}.
		\bjtitle{Chemistry of Materials}
		\bvolume{37}(\bissue{3}),
		\bfpage{1160}--\blpage{1173}
		(\byear{2025})
	\end{barticle}
	\endbibitem

	\bibitem[\protect\citeauthoryear{Matsui et~al.}{2025}]{matsui2025tailoring}
	\begin{barticle}
		\bauthor{\bsnm{Matsui}, \binits{N.}},
		\bauthor{\bsnm{Mori}, \binits{K.}},
		\bauthor{\bsnm{Saito}, \binits{T.}},
		\bauthor{\bsnm{Noi}, \binits{K.}},
		\bauthor{\bsnm{Fujinami}, \binits{S.}},
		\bauthor{\bsnm{Park}, \binits{Y.}},
		\bauthor{\bsnm{Tojigamori}, \binits{T.}},
		\bauthor{\bsnm{Suzuki}, \binits{K.}},
		\bauthor{\bsnm{Abe}, \binits{T.}},
		\bauthor{\bsnm{Kanno}, \binits{R.}}:
		\batitle{Tailoring polyhedral linkage in K$_{1-x}$Ba$_x$M$_2$F$_{7+x}$ (M~=~Yb, Lu) for enhanced fluoride-ion conductivity}.
		\bjtitle{Chemistry of Materials}
		\bvolume{37}(\bissue{13}),
		\bfpage{4798}--\blpage{4806}
		(\byear{2025})
	\end{barticle}
	\endbibitem

	\bibitem[\protect\citeauthoryear{Ward et~al.}{2018}]{ward2018matminer}
	\begin{barticle}
		\bauthor{\bsnm{Ward}, \binits{L.}},
		\bauthor{\bsnm{Dunn}, \binits{A.}},
		\bauthor{\bsnm{Faghaninia}, \binits{A.}},
		\bauthor{\bsnm{Zimmermann}, \binits{N.E.R.}},
		\bauthor{\bsnm{Bajaj}, \binits{S.}},
		\bauthor{\bsnm{Wang}, \binits{Q.}},
		\bauthor{\bsnm{Montoya}, \binits{J.}},
		\bauthor{\bsnm{Chen}, \binits{J.}},
		\bauthor{\bsnm{Bystrom}, \binits{K.}},
		\bauthor{\bsnm{Dylla}, \binits{M.}}
		et~al.:
		\batitle{Matminer: An open source toolkit for materials data mining}.
		\bjtitle{Computational Materials Science}
		\bvolume{152},
		\bfpage{60}--\blpage{69}
		(\byear{2018})
	\end{barticle}
	\endbibitem



\end{thebibliography}

\end{document}